\documentclass[12pt]{iopart}

\usepackage{iopams}
\usepackage{amssymb,latexsym,amsthm}
\usepackage{graphicx}
\usepackage{subcaption}
\usepackage{palatino, url, multicol}
\usepackage{placeins}
\usepackage{xcolor}
\usepackage{colordvi}

\begin{document}

\title[Gravitational Entropy in Szekeres Class I Models]{Gravitational Entropy in Szekeres Class I Models}

\author{Fernando A. Piza\~na$^{1,\star}$, Roberto A. Sussman$^{1,\dagger}$ and Juan Carlos Hidalgo$^{2,\S}$}

\address{$^1$Instituto de Ciencias Nucleares, Universidad Nacional Aut\'onoma de M\'exico (ICN-UNAM), 04510 Ciudad de M\'exico, M\'exico. \\
$^2$Instituto de Ciencias F\'isicas, Universidad Nacional Aut\'onoma de M\'exico (ICF-UNAM), 62210 Cuernavaca, Morelos, M\'exico.}
\ead{$^\star$klesto92@ciencias.unam.mx, $^\dagger$sussman@nucleares.unam.mx, $^\S$hidalgo@icf.unam.mx}
\vspace{10pt}

\begin{abstract}
Developing a self-consistent notion of gravitational entropy in the context of cosmological structure formation has been so far an elusive task. Various theoretical proposals have been presented, initially based on Penrose's Weyl Curvature Hypothesis, and variations of it. A more recent proposal by Clifton, Ellis, and Tavakol (CET) considered a novel approach by defining such entropy from a Gibbs equation constructed from an effective stress-energy tensor that emerges from the 'square root' algebraic decomposition  of the Bel-Robinson tensor, the simplest divergence-less tensor related to the Weyl tensor. Since, so far all gravitational entropy proposals have been applied to highly restrictive and symmetric spacetimes, we probe in this paper the CET proposal for a class of much less idealized spactimes (the Szekeres class I models) capable of describing the joint evolution of arrays of arbitrary number of structures: overdensities and voids, all placed on selected spatial locations in an asymptotic  $\Lambda$CDM backgound. By using suitable covariant variables and their fluctuations, we find the necessary and sufficient conditions for a positive CET entropy production to be a negative sign of the product of the density and Hubble expansion fluctuations. To examine the viability of this theoretical result we examine numerically the CET entropy production for two elongated over dense regions surrounding a central spheroidal void, all evolving jointly from initial linear perturbations at the last scattering era into present day Mpc-size CDM structures. We show that CET entropy production is positive for all times after last scattering at the precise spatial locations where structure growth occurs and where the exact density growing mode is dominant. The present paper provides the least idealized (and most physically robust) probe of a gravitational entropy proposal in the context of structure formation.
\end{abstract}

\bibliographystyle{unsrt}

%
\vspace{2pc}
\submitto{\CQG}
%
%
%

\section{Introduction}
A self-consistent  formulation of gravitational entropy in the context of large scale structure formation in a cosmological scenario is still an open and appealing concept that would enhance our understanding of self-gravitating systems in a cosmological scale. This entropy must provide theoretical elements for relating the gravitational interaction with an ``arrow of time'', meaning an irreversible temporal direction of structure formation ({\it i.e.} growth of inhomogeneity or ``lumpiness'') associated with physical observers. Evidently, being a separate frame-dependent concept, it is not expected nor required that this gravitational entropy should coincide with its thermodynamic counterparts associated with the field sources and/or the holographic Hawking-Beckenstein formulation for black holes and localized system in asymptotically flat space-times. However, a correspondence with the holographic formulation must exist in the appropriate limits.

The first proposal for a gravitational entropy (distinct from the holographic one) was the ``Weyl Curvature Hypothesis'' suggested by Roger Penrose in terms of invariant (frame independent) scalars constructed with the Weyl tensor (see \cite{Pelavas:2004ih} and its cited references). Penrose argued that considering the Weyl tensor was a necessary consequence of the fact that it does not vanish in the absence of field sources, and thus it can be associated with a ``free'' sourceless gravitational field. There is an extensive literature applying the ideas of Penrose to different spacetimes and contexts, including an extension of Penrose's original proposal which considers different contractions of the Weyl tensor and ratios of these scalars to contractions of the Ricci and Riemann tensors (see \cite{Guha:2019fun, Zhao:2017kmk, Gregoris:2021fiw, Malik:2019kvh, Li:2014ffa, Mishra:2014mna}).

An important limitation of the gravitational entropies based on Penrose approach was reported by Bonnor \cite{bonnor1987arrow}. By looking at a class of collapsing heat conducting and energy radiating spheres, he found that the ``arrow of time'' behaves in the opposite direction to that expected from a gravitational entropy: the latter decreased as the spheres collapsed. Further development was furnished by Pelavas and Lake \cite{Pelavas:1998dk} who considered spacetimes compatible with homothetic symmetries to argue that a viable gravitational entropy should not be defined in terms of curvature invariants. Pelavas and Coley \cite{Pelavas:2004ih} continued this work by defining gravitational entropy in terms of  more general dimensionless scalar functions, including one given in terms of the Bel-Robinson tensor, applying these functions to Szekeres class II and Bianchi type $VI_{h}$ models and showing that all their definitions yield a monotonic increase of gravitational entropy that approaches an asymptotic equilibrium saturation stage. 

Rudjord and Gr{\o}n \cite{Rudjord:2006vj} considered Weyl tensor invariants and the cosmological constant as elements of a ``curvature conjecture'' to describe the holographic entropy of black holes and horizons in terms of the Weyl tensor. They introduced the notion of the ``total entropy of a gravitational system'' by the sum of its thermodynamical part (associated to the field sources) and its geometric part (associated to the Weyl curvature). By applying this conceptual notion to Schwarzschild-de Sitter and de Sitter spacetimes, they found that the entropy of the cosmological horizon is either not geometrical in origin or might depend on a thermodynamical interpretation of the cosmological constant; thus refuting the notion of an holographic entropy arising from a Weyl tensor related entropy of the free gravitational field. The results of these papers suggested the need for a more robust formulation of a non-holographic notion of gravitational entropy. 

Clifton, Ellis and Tavakol (CET) \cite{Clifton:2013dha} modified Penrose's original proposal (and its extensions) by arguing that gravitational entropy must be related to the Weyl tensor, but need not be an invariant, frame-independent scalar. Rather, CET argued that the gravitational entropy must be obtained for specific observer congruences through a Gibbs 1-form that follows from an effective stress-energy tensor $\mathcal{T}_{gr}^{ab}$ constructed by the algebraic `square root' decomposition of the 4th-order Bel-Robinson tensor $T_{abcd}$, the only divergence-less tensor related to the Weyl tensor. In this proposal, the gravitational entropy ${s}_{gr}$ depends on the 4-velocity  field $u^a$ of the sources, and is thus a local pointwise quantity, with the effective energy-momentum tensor $\mathcal{T}_{gr}^{ab}$  representing a ``geometric fluid'', completely unrelated to the energy momentum tensor of the field sources.

The CET entropy proposal must satisfy the following properties expected from an entropy formalism: 
\begin{itemize}
\item it is globally a constant ($\dot s_{gr}=0$) for conformally flat spacetimes, including Friedmann-Lema\^itre-Robertson-Walker (FLRW) models
 (since $C_{abcd}=0$), 
\item it must provide a measure of the increase of lumpiness (local anisotropy and inhomogeneity) along the simultaneity surfaces of cosmological observers, 
\item it must reduce to the Bekenstein-Hawking holographic entropy when applied to a Schwarzschild black hole in the static frame, 
\item it must increase monotonically along $u^a$ for non-empty spacetimes as structure formation occurs, 
\item it must increase when applied to perturbations on an FLRW background that evolve through their growing mode.  
\end{itemize}
If these conditions hold, CET conclude that ${s}_{gr}$ provides a temporal direction for configurations that initiate as linear perturbations, eventually growing into fully non-linear structures. The main problem of this proposal is that the irreducible `square root' decomposition of the Bel-Robinson tensor is unique only for Petrov D and N spacetimes (Coulombian and Wavelike spacetimes). Evidently, this is still an unresolved limitation of this formalism. Let us now provide a summary of previous work on gravitational entropy. 

Bolejko and Stoeger \cite{Bolejko:2013doa} looked at the evolution of special cases of spherically symmetric inhomogeneous cosmological models undergoing an intermediate homogeneization stage, which might be identified with a primordial inflationary stage (assuming a viscous fluid source) and for a late Universe (assuming a dust source) with the spatial curvature being negligible compared with the Ricci curvature (similar conditions occur as saddle points in dynamical system studies of Bianchi models). In both cases of intermediate homogeneization, where the decaying mode is dominant, the gravitational entropy decreases in various suggested definitions, but also the decaying mode is dominant. However, these models are far from generic, and in the case of the early Universe the gravitational entropy ends up growing in the late post-inflationary stage as the growing mode becomes dominant. 

The CET proposal has been examined for spherically symmetric Lema\^\i tre-Tolman-Bondi (LTB) dust models by  Sussman and Larena \cite{Sussman:2013xpa}, looking at generic models and the specific case of single local cosmic voids in an FLRW background (with both zero and positive $\Lambda$)  \cite{Sussman:2015bea}. In these papers the CET entropy is studied by means of covariant quasilocal variables: weighted average functions of covariant variables, with local variables defined as fluctuations around them. For generic models, the necessary and sufficient conditions for a positive entropy production occur when the product of the density and Hubble scalar fluctuations is negative, which occurs when the exact density growing mode \cite{Sussman:2013qya} is dominant. As we show further ahead in the present paper, this happens also with Szekeres models of class I that generalize LTB models. In the first paper \cite{Sussman:2013xpa} the authors show that the Hosoya-Buchert-Morita (HBM) \cite{Hosoya:2004nh} entropy proposal based on Information Theory provides the same outcomes as the CET proposal, but given as a non-local domain dependent quantities. In the second paper, the authors show that the CET entropy diverges asymptotically as a logarithm of cosmic time if $\Lambda=0$ holds in the FLRW background, while it converges to a saturation value proportional to an asymptotic gravitational temperature $\propto \sqrt{\Lambda}$ for a background dominated by Cold Dark Matter and a cosmological constant ($\Lambda$CDM). The results of Sussman and Larena coincide with (and provides an exact generalization to) the analysis by Marozzi, Uzan, Umeh and Clarkson \cite{Marozzi:2015qra} based on linear dust perturbations on an FLRW background with zero decaying mode.  

In \cite{Perez:2014mba}, P\'erez and Romero showed that the CET formalism applied to a Kerr black hole metric does not reproduce the Bekenstein-Hawking entropy for the specific frame they consider, showing further that the scalar formed by the ratio of the Weyl scalar to the Kretschmann scalar is more suitable for this spacetime. A study of the CET entropy on black holes undertaken by Acquaviva, Ellis, Doswami and Hamid \cite{Acquaviva:2014owa} analyzed the thermodynamic properties of a Schwarzschild  black hole formed through the Oppenheimer-Snyder-Datt collapse, proving that the change in gravitational entropy outside the collapsing dust sphere is related to the variation of its surface area, which allows them to relate the CET entropy to the Bekenstein-Hawking entropy of the resulting black hole. Chakraborty, Guha and Goswami \cite{Chakraborty:2019mqv} applied the CET entropy for different anisotropic cosmologies (including the quasi-flat Szekeres class II model) by looking at the Penrose Weyl curvature hypothesis around the initial singularity and identifying for which models (and under which assumptions) the Weyl hypothesis is valid. Chakraborty and colleagues \cite{Chakraborty:2021uat} applied the entropy definition from Rudjord and Gr{\o}n \cite{Rudjord:2006vj}, as well as the CET proposal, to traversable wormholes (though they only considered the CET proposal for static spacetimes). Villalba, Bargue\~no, Vargas and Contreras \cite{Villalba:2020dbv} analyze the connections between the Newmann-Penrose scalars with the equation of state of asymptotically Anti-de Sitter Reissner-N\"ordstrom black holes using CET's Bel-Robinson's square root thermodynamical description proposing conditions at the horizons in terms of pressures and energy densities as alternative thermodynamical definitions of these surfaces. 

Finally, considering less ``orthodox'' approaches, I V. Ruchin, O. Vacaru and S. I. Vacaru \cite{Ruchin:2013azz} define a {\it sui generis} W-entropy for gravitational fields by means of F- and W- functionals, obtaining a statistical thermodynamic description of gravitational interactions analogous to that of CET. Acquaviva, Kofro\v{n} and Scholtz \cite{Acquaviva:2018mqb} showed that the 'square root' of the Bel-Robinson  tensor leads to an effective energy-momentum tensor with  dissipative terms, while in the same context Goswami and Ellis \cite{Goswami:2018zfo} examined the energy and momentum transfer associated with these terms for both Petrov types D and N.  Considering more ``exotic'' sources, Gregoris, Ong and Wang \cite{Gregoris:2020usv} provide a counter-example to Penrose's Weyl Curvature Hypothesis through inhomogeneous models associated with chameleon massless scalar field with anisotropic shearing effects. For these sources the CET entropy increases while the Weyl curvature decreases due to the growth of the shear scalar, thus inferring a possible connection with the formation of primordial structures such as the Large Quasar Groups. 

It is important to note (as shown above) that most work in probing the viability of the CET gravitational entropy has been carried out in highly idealized and symmetric spacetimes, including LTB (and other) models with spherical symmetry that allow for the description of a single structure in an FLRW background. It is thus a natural extension of previous work to overcome the constraints of spherical symmetry (and other isometry groups) by probing this proposal on a class of much less idealized models that do not admit isometries and allow for the description of arrays of arbitrary number of jointly evolving structures. The natural generalization of LTB models that complies with these desired properties is the class of quasi-spherical Szekeres Class I models, for which, as we show throughout the paper, the CET entropy does provide a ``time arrow'' for an irreversible growth of gravitational entropy tied with the growth of several CDM structures in a $\Lambda$CDM background in conditions in which the exact density growing mode is dominant.

This paper is organized as follows: in \sref{sec:SzekI} we introduce the Szekeres Class I models in FLRW-like coordinates, showing how  Einstein's field equations reduce to a first-order system in terms of quasilocal scalars and non-spherical fluctuations. We examine the connection between the exact dynamics of Szekeres models with that of gauge-invariant cosmological perturbations by comparing in \sref{sec:FvsC} the notion of density contrast with an analogous exact quasilocal density fluctuation. In \sref{sec:CETSzekI} we outline the CET gravitational entropy formalism for Szekeres Class I models, finding the necessary and sufficient conditions for a positive entropy production. A generic numerical example describing a central spheroidal void surrounded by two over-densities (in a $\Lambda$CDM background) is presented and developed in full in \sref{sec:NumEx}. This example shows how the CET entropy grows precisely in the regions where these structures grow. In \sref{sec:Disc} we provide an analytic guideline in order to relate our results to those obtained previously with LTB models. In the same section we provide an elaborate a discussion of our results, showing that the conditions for positive entropy production correspond to the exact generalization of a dominant growing mode, with the CET gravitational entropy when $\Lambda>0$ evolving towards asymptotic values proportional to an asymptotic gravitational temperature. 

Finally, we provide three appendicess: \ref{sec:coords} deals with the Szekeres I metric in its usual coordinates and the transformation to stereographic coordinates. \ref{sec:IntCond} outlines the integrability conditions of the Gibbs 1-form. \ref{analytic} discusses the analytic solutions of the Friedman-like equations for models with $\Lambda = 0$.

\section{\label{sec:SzekI}Quasi-spherical Szekeres Models of Class I}

Szekeres models are exact solutions of Einstein's equations that (in general) do not admit isometries. They are characterized  by a dust source with a positive cosmological constant: $T_{ab}=\rho u_au_b-\Lambda g_{ab}$ in a comoving frame $u^a =\delta^a_0$. The solutions are split in two classes (I and II), each containing symmetric subcases, such as axial symmetry, as well as subcases admitting 3-dimensional isometry groups acting on 2-dimensional orbits: spherical, plane or pseudo-spherical symmetry. Szekeres models (of both classes) in which such limiting case is spherically symmetric are known as ``quasi-spherical'' models (see terminology and full classification in \cite{Krasinski:1997yxj}). In the present paper we will deal only with quasi-spherical models of class I whose spherically symmetric limit is the class of dust Lema\^\i tre-Tolman-Bondi (LTB) models. To simplify the terminology we will use the term ``Szekeres I models'' to denote these models.

The models can be represented by various coordinate systems (see \sref{sec:NumEx} and \ref{sec:coords}). A simple and useful representation is furnished by the following diagonal FLRW-like line element \cite{Sussman:2011bp}:
\begin{equation}
    \rmd s^2 = - \rmd t^2 +a^2\left[\frac{\mathcal{G}^2\rmd r^2}{1- K_{qi}r^2}+\frac{r^2}{\mathcal{E}^2}(\rmd x^2 +  \rmd y^2)\right]\label{eq:FLRWlike},
\end{equation}
where $a=a(t,r)$,\,\,$K_{qi} = K_{qi}(r)$ (see the interpretation of this free function in equation (\ref{eq:sclaws})), with the subindex ${}_i$ denoting henceforth evaluation of any scalar function at an arbitrary initial time slice. The remaining functions are 
\begin{eqnarray}
    \fl \mathcal{G}=\Gamma - \mathbf{W}, \quad \Gamma = 1+\frac{ra'}{a},\quad \mathbf{W}=\frac{r\mathcal{E}'}{\mathcal{E}},\quad \mathcal{E} = \frac{S}{2}\left[1+\left(\frac{x-P}{S}\right)^2+\left(\frac{y-Q}{S}\right)^2\right],\label{eq:GammaEdefs} 
\end{eqnarray}
with $a'=\partial a/\partial r$ and $S=S(r),\,P=P(r),\,Q=Q(r)$ denoting the dipole parameters.  This form of the metric admits the following orthonormal tetrad
\begin{equation}
    u_a=-\delta_a^0, \quad \mathbf{r}_a = \frac{a\mathcal{G}}{\sqrt{1- K_{qi}r^2}}\delta_a^r, \quad \mathbf{x}_a = \frac{ar}{\mathcal{E}} \delta_a^x, \quad \mathbf{y}_a = \frac{ar}{\mathcal{E}} \delta_a^y.\label{eq:tetrad}
\end{equation}
In a comoving frame, where $u^a=\delta^a_0$ with $h_{ab}=u_au_b+g_{ab}=a^2\gamma_{ij}\delta_a^i\delta_b^j$, Szekeres I models are characterized by the following covariant scalars \cite{Krasinski:1997yxj,Sussman:2011bp}
\begin{eqnarray}
\eqalign{\rho= u_a u_b T^{ab}\,\, \hbox{(density)},\quad K=\frac16\,{}^{(3)}\mathcal{R}\,\, \hbox{(spatial curvature)},\\ H =\frac{\Theta}{3}=\frac13 u^a_{;a} \,\, \hbox{(Hubble scalar)}},\label{eq:covscals}
\end{eqnarray}
where $\Theta = u^a\,_{;a}$ is the expansion scalar, plus the spacelike traceless symmetric shear and electric tensors $\sigma_{ab}=u_{(a;b)}-\Theta/3\,h_{ab}$ and $E_{ac}=u^b u^d C_{abcd}$, where $C_{abcd}$ is the Weyl tensor. These covariant objects can be expressed in the following concise and elegant form
\footnote{In previous articles on gravitational entropy in LTB models (for example \cite{Sussman:2013xpa,Sussman:2015bea}) we used the notation $\mathbf{D}^A$ or $\mathbf{D}_q(A)$ for the fluctuations $A-A_q$. In the present article these LTB fluctuations will be denoted as $\mathbf{d}_q^A$, see equation \eref{Ddrelation} in \sref{sec:Disc}} further ahead.
\begin{eqnarray}
A &=& A_q+\mathbf{D}_q^A=A_q+\frac{rA'_q}{3\mathcal{G}}, \quad \hbox{for} \quad A=\rho,\,K,\,H,\label{eq:locscals}\\
 &\Longrightarrow&\quad  \mathbf{D}_q^A= A-A_q = \frac{rA'_q}{3\mathcal{G}},\label{eq:flucts}\\
\sigma_{ab} &=& -\mathbf{D}_q^{H}\,\xi_{ab},\qquad E_{ab}=-\frac{4\pi}{3}\mathbf{D}_q^\rho\,\xi_{ab}=\Psi_2\,\xi_{ab},\label{eq:shearEWeyl}
\end{eqnarray}
where $\Psi_2$ is the conformal invariant of Petrov type D spacetimes and $\xi^a_b=-2\mathbf{r}^a\mathbf{r}_b+\mathbf{x}^a\mathbf{x}_b+\mathbf{y}^a\mathbf{y}_b= \hbox{diag}[0,-2,1,1]$ is the common trace-less and divergence-less eigenframe ($\xi_a^b\,_{;b}=0$). 

As shown in previous work \cite{Sussman:2011bp,Sussman:2015fwa,Sussman:2015wna}, it is very useful to describe the dynamics of Szekeres I models in terms of quasilocal scalars (to be denoted as \textit{q-scalars}) $A_q = A_q(t,r)$ associated to local covariant scalars in (\ref{eq:covscals}) $A=\rho,\,K,\,H$ by means of the following weighted average functions defined by 3-dimensional integrals along the rest frames:
\footnote{The q-scalars were originally derived for LTB models and its definition was extended to Szkeres I models in \cite{Sussman:2011bp}. Strictly speaking, they are distributions, given as functions of the domain boundary marked by the ``radial'' coordinate $r$, constructed from the weighed average functional with correspondence rule (\ref{eq:Aqdef}). See comprehensive discussion in \cite{Sussman:2014cda} and in \cite{Sussman:2013yq} for LTB models.}
\begin{eqnarray}
A_q =\frac{\int_V{A\,F\,\rmd \mathcal{V}}}{\int_V{F\,\rmd \mathcal{V}}}=\frac{\int_0^ r{\int_{-\infty}^\infty{ \int_{-\infty}^{\infty}{ A\,F\,\mathcal{J} \rmd r \rmd x \rmd y}}}}{\int_0^ r{\int_{-\infty}^\infty{ \int_{-\infty}^{\infty}{ F\,\mathcal{J} \rmd r \rmd x \rmd y}}}},\label{eq:Aqdef}\\
F\equiv \sqrt{1- \mathcal{K}_{qi}r^2},\qquad \mathcal{J}=\sqrt{\hbox{det}(h_{ab})}=\frac{a^3\,r^2\,\mathcal{G}}{F\mathcal{E}^2},\label{eq:GFJdefs}
\end{eqnarray}
which, from \eref{eq:Aqdef}, satisfy the following scaling laws with respect to initial values at an arbitrary time slice $t=t_i$ 
\begin{equation}
\rho_q=\frac{\rho_{qi}}{a^3},\qquad K_q=\frac{K_{qi}}{a^2},\qquad H_q = \frac{\dot a}{a}.\label{eq:sclaws}
\end{equation}
Szekeres I models contain the spherically symmetric LTB dust models in the limit  $S'=P'=Q'=0\,\Rightarrow\,\, \mathcal{E}$ = const., after performing a trivial transformation to spherical coordinates. We will denote by ``LTB seed model'' the LTB model obtained in this limit for any given Szekeres I model. Notice that the q-scalars are independent of the non-spherical coordinates $x,y$, hence they are common functions with the LTB seed model. Consequently, the deviation from spherical symmetry in Szekeres I models is encoded in the fluctuations $\mathbf{D}_A$ given in equations \eref{eq:locscals}-\eref{eq:shearEWeyl}.   

Einstein's equations with the variables we have defined become the following fully covariant, first-order system of evolution equations and constraints \footnote{To solve the system numerically these variables will be suitably normalized to render them dimensionless.}
\begin{eqnarray}
    \dot{\rho}_q &=& -3\rho_q H_q,\label{eq:rhoev} \\
    \dot{H}_q &=& -H_q^2-\frac{4\pi}{3}\rho_q+\frac{8\pi}{3}\Lambda,\label{eq:Hqev}\\
    \dot{\mathbf{D}_q^{\rho}}&=&-3(H_q+\mathbf{D}_q^{H})\mathbf{D}_q^{\rho}-3\rho_q\mathbf{D}_q^H,\label{eq:Drhoev}\\
    \dot{\mathbf{D}}_q^H &=& -(2H_q+3\mathbf{D}_q^H)\mathbf{D}_q^H-\frac{4\pi}{3}\mathbf{D}_q^{\rho},\label{eq:Dhev}\\
    \dot{a} &=& a\,H_q,\label{eq:aev}\\
    \dot{\tilde{\mathcal{G}}} &=& 3\tilde{\mathcal{G}}\mathbf{D}^H, \quad \tilde{\mathcal{G}} = \frac{\mathcal{G}}{1-\mathbf{W}} =  \frac{\Gamma-\mathbf{W}}{1-\mathbf{W}}\label{eq:Gev},
\end{eqnarray}
together with the algebraic constraints
\begin{eqnarray}
    H_q^2=\frac{8\pi}{3}(\rho_q+\Lambda)-K_q,\label{eq:Hconstr}\\
    2H_q\mathbf{D}_q^H = \frac{8\pi}{3}\mathbf{D}_q^{\rho}-\mathbf{D}_q^{K}.\label{eq:DHconstr}
\end{eqnarray}
To solve the system \eref{eq:rhoev}-\eref{eq:DHconstr} we need to specify initial conditions at an arbitrary time slice $t=t_i$, where we are assuming (as part of the freedom to rescale spatial coordinates) that $a_i=\Gamma_i=\tilde{\mathcal{G}}_i=1$ and that all radial gradients vanish at $r=0$ (a special worldline along which the shear and electric Weyl tensors vanish which is analogous to the regular center of symmetry in a spherically symmetric spacetimes and it corresponds to an observer for which local observations are isotropic). Since the constraints are preserved at all times, then it is sufficient to prescribe as initial conditions any two of the three quasilocal free parameters $\{\rho_{qi},\,K_{qi},\,H_{qi}\}$, plus the three free parameters $\{S,\,P,\,Q\}$ of the dipole $\mathcal{E}$. Since $\mathbf{W}$ is independent of $t$, both $\tilde{\mathcal{G}}$ and $\mathcal{G}$ satisfy the same evolution equation. Any two of the initial fluctuations can be obtained from \eref{eq:locscals} and \eref{eq:DHconstr} applied at $t_i$, namely $\mathbf{D}^A_i= rA'_{qi}/[3(1-\mathbf{W})]$ for $A=\rho,\,K,\,H$.    

\section{\label{sec:FvsC}Connection to cosmological perturbations: density fluctuation vs. density contrast}

The quasi-local average functions and their fluctuations are fully covariant  variables that will be useful to apply the CET gravitational entropy formalism to Szekeres models. In spite of this, the most common variables employed for the description of cosmological phenomena are the gauge-invariant perturbations derived from the metric \cite{Malik:2008yp, Malik:2008im}. However, the dynamics of Szekeres models described by the q-scalars and their fluctuation can be directly related to the dynamics of the models in terms of gauge invariant quantities in the isochronous comoving gauge. While in the linear regime the dynamical equations for Szekeres models in both formalisms become identical, there are important conceptual difference some of which we discuss below (see extensive discussion in \cite{Sussman:2017otk}).   

The gauge-invariant theory of cosmological linear perturbations relies on quantities constructed as ``contrasts'' with respect to quantities in an FLRW background. These are simple quotients of relative differences between a ``perturbed'' value and its FLRW equivalent, a notion that can be formally generalized to the non-linear regime, but a distinct notion to the quasilocal fluctuations used in this paper, which are exact quantities (or ``exact perturbations'') denoting a difference with a weighted averaged function. 

In previous work we have shown \cite{Sussman:2017otk} that the dynamics of Szekeres dust models as described here in terms of ``exact perturbations'' can also be related to quantities defined as ``contrasts'', either in the linear regime or as exact quantities. For example, assuming the existence of an asymptotic FLRW background as $r\to\infty$, we have $\rho_q\to \bar\rho$ in this limit \cite{Sussman:2017otk} (we denote henceforth all quantities defined in the FLRW asymptotic with an overbar), the usual density contrast in gauge invariant cosmological perturbations, $\delta$ (denoted by $\Delta_{\hbox{\tiny{(as)}}}^\rho$ in \cite{Sussman:2017otk}), can be compared with an analogous dimensionless exact fluctuation $\Delta_q^\rho$
\begin{equation}
    \delta= \frac{\rho-\bar\rho}{\bar \rho} \qquad \Delta_q^\rho=\frac{\mathbf{D}_q^\rho}{\rho_q}=\frac{\rho-\rho_q}{\rho_q} 
\end{equation}
whose relation is given by
\begin{equation}
    \delta=\Delta_q^{\rho}\,\frac{\rho_q}{\bar{\rho}}-  \left(1- \frac{\rho_q}{\bar{\rho}} \right).
\end{equation}
As shown in \cite{Sussman:2017otk}, in the linear regime (either close to the FLRW background as $r\to\infty$ or  close to last scattering taken as initial time slice) both types of variables are equivalent, since we have $\rho_q\approx \rho\approx \bar\rho$ and $H_q\approx H\approx \bar H$, so that $\delta \approx \Delta_q^\rho$. 

However, in a fully non-linear regime (small redshifts and/or far from the asymptotic background) that requires treatments with non-perturbative variables, $\delta$ and $\Delta_q^\rho$  can differ significantly: both $\delta$ and $\Delta_q^\rho$ taken as exact quantities satisfy distinct non-linear second order evolution equations \cite{Sussman:2017otk}:
\begin{eqnarray}
&{}& \ddot\delta-\frac{\dot\delta^2}{1+\delta}+\left[6\bar{H}-4H_q\right]\dot \delta-\left[4\pi\bar\rho\,\delta-18(\bar{H}-H_q)^2\right](1+\delta)=0, \label{perturb2ord1}\\
&{}& \ddot \Delta_q^\rho-\frac{2[\dot\Delta_q^\rho]^2}{1+\Delta_q^\rho}+2H_q\dot\Delta_q^\rho-4\pi\rho_q\Delta_q^\rho\,[1+\Delta_q^\rho]=0,\label{perturb2ord2} 
\end{eqnarray}
with both reducing to the familiar linear dust perturbation in the synchronous gauge: $\ddot\delta +2\bar{H}\,\dot\delta-4\pi\bar\rho\,\delta=0$.

Note also that there is a difference in the physical meaning of the sign in both functions $\delta$ and $\Delta_q^\rho$. The sign of $\delta$ provides for all times a simple straightforward distinction between over-densities ($\delta>0$: local density larger than background density) and voids ($-1<\delta<0$ local positive density less than background density).  For spherically symmetric LTB models, describing a single central structure in an FLRW background, the sign of the spherical equivalent of  $\Delta_q^\rho$ is also simple: it only depends on the sign of the gradient $\rho'_q$ (assuming absence of shell crossings so that $\Gamma>0$), hence the signs of $\Delta^\rho_q$ are exactly the opposite of those of $\delta$: it is negative for central over-densities ($\rho'_q<0$ for $r$ increasing away from the center of symmetry at $r=0$) and it is positive for density voids for the opposite reason. 

On the other hand, for Szekeres I models the sign of $\Delta_q^\rho$ (and thus $\mathbf{D}_q^\rho$) becomes more subtle: it is like in LTB models for a central spheroidal structure enclosing $r=0$  (over-density or void), but this sign does not necessarily distinguish in a clear cut way the concavity or convexity of $\rho$ (overdensities or voids) in structures that are away from the central spheroidal region, as such structures can arise for various combinations of signs of $\rho'_q$ and also depend on the gradients of free parameters of the Szekeres dipole $\mathbf{W}$ in \eref{eq:GammaEdefs}. An example of the differences of Szekeres I radial profiles with typical void or overdensity profiles in spherical symmetry emerge in the graphs of initial functions of the numerical example displayed in figure \ref{Fig:perfini} (see comprehensive discussion of concavity/convexity of Szkeres I scalars in \cite{Sussman:2015fwa,Sussman:2015wna}).  
        
It is important to clarify this difference, since (as we show in sections \sref{sec:SdotPlus} and \sref{section:numerical} further ahead) the sign of the product $\mathbf{D}_q^\rho \mathbf{D}_q^H$ is directly related to that of the CET entropy growth, irrespective of whether the structures are overdensities or voids.

\section{\label{sec:CETSzekI} CET entropy for Szekeres I models}

In \cite{Clifton:2013dha} CET presents a gravitational entropy proposal for space-times which are Petrov type D and N. This formalism is based on the irreducible algebraic decomposition of the Bel-Robinson tensor into two rank-2 tensors developed by Bonilla and Senovilla in \cite{Bonilla:1997ink}, for the particular cases of Petrov D and N, these two tensors are equal and we call it the \emph{square root} of the Bel-Robinson tensor. This \emph{square root} for the Szekeres I models (which are Petrov D space-times) is given in terms of the Newman-Penrose (NP) tetrad $l_a$, $k_a$, $m_a$ and $\bar{m}_a$ and the NP scalar $\Psi_2$ by
\begin{equation}
t_{ab} = 3 |\Psi_2|(m_{(a}\bar{m}_{b)}+l_{(a}k_{b)})+fg_{ab}, \quad \mbox{with} \quad f=-\frac{|\Psi_2|}{2},\label{eq:tabdef}
\end{equation}
leading to the ``effective'' energy momentum tensor of the following geometric fluid 
\begin{equation}
8 \pi \mathcal{T}_{ab}^{\textrm{\tiny{(gr)}}} = \alpha t_{ab}=|\Psi_2| (\mathbf{x}_a\mathbf{x}_b+\mathbf{y}_a\mathbf{y}_b-2(-u_au_b+\mathbf{r}_a\mathbf{r}_b)),\label{cetTab}
\end{equation}
where $\alpha>0$ is a constant of $\sim O(1)$ and $\{\mathbf{x}_a,\,\mathbf{y}_a,\,\mathbf{r}_a,\,u_a\}$ are the orthonormal tetrad \eref{eq:tetrad} associated with the NP tetrad in \eref{eq:tabdef}. Note that the effective tensor meets the conservation equations $\mathcal{T}_{\textrm{\tiny{(gr)}}}^{ab}\,_{;b} = 0$. From this effective tensor CET obtain the gravitational thermodynamic variables
\begin{equation}
\eqalign{8\pi \rho_{\textrm{\tiny{(gr)}}}=8\pi \mathcal{T}^{\textrm{\tiny{(gr)}}}_{ab}u^au^b = 2\alpha |\Psi_2|, \quad 8\pi p_{\textrm{\tiny{(gr)}}}=8\pi \frac13h^{ab}\mathcal{T}^{\textrm{\tiny{(gr)}}}_{ab}= 0, \\
 8\pi  \Pi^{\textrm{\tiny{(gr)}}}_{ab}=8\pi \left[h_{(a}^ch_{b)}^d-\frac13 h_{ab}h^{cd}\right]\mathcal{T}^{\textrm{\tiny{(gr)}}}_{cd}= \alpha |\Psi_2|(-2\,\mathbf{r}_a \mathbf{r}_b+\mathbf{x}_a \mathbf{x}_b+\mathbf{y}_a\mathbf{y}_b),\\ 8\pi  q^{\textrm{\tiny{(gr)}}}_a=-8\pi u^c\mathcal{T}^{\textrm{\tiny{(gr)}}}_{cb}h^b_a=0,}
\end{equation}
which, using $(u_a\mathcal{T}_{\textrm{\tiny{(gr)}}}^{ab})\,_{;b}=\mathcal{T}_{\textrm{\tiny{(gr)}}}^{ab}\,u_{a;b}$ and the decomposition $u_{a;b} = -\dot{u}_au_b+\frac{1}{3}\Theta h_{ab} + \sigma_{ab} +\omega_{ab}$ (where $\dot u_a =\omega_{ab}=0$ for Szkeres I models) CET obtain the following Gibbs-like equation 
\begin{equation}
    T_{\textrm{\tiny{(gr)}}}\dot{s}_{\textrm{\tiny{(gr)}}} = \partial_t(\rho_{\textrm{\tiny{(gr)}}} V),\label{eq:gibbs1}
\end{equation}
where $V=F\mathcal{E}\mathcal{J}/r^2 = a^3\,\mathcal{G}$ is the volume defined by the expansion scalar $\Theta = \dot{V}/V =\dot{\mathcal{J}}/\mathcal{J}$ and $T_{\textrm{\tiny{(gr)}}}$ is an integration factor of the Gibbs-like equation that must be interpreted as the gravitational temperature associated to the gravitational entropy $s_{\textrm{\tiny{(gr)}}}$. Given the lack of information on this gravitational temperature in the Bel-Robinson tensor, CET propose the following expression
\begin{equation}
    T_{\textrm{\tiny{(gr)}}}=\frac{|\nabla_bu_ak^al^b|}{\pi}=\frac{|\dot{u}_a\,\mathbf{r}^a+H+\sigma_{ab}\mathbf{r}^a\mathbf{r}^b|}{2\pi}=\frac{|H-2\sigma|}{2\pi}.
\end{equation}
which can be motivated physically to connect with thermodynamic considerations applicable to a wide variety of spacetimes.    
 
As shown in \eref{eq:shearEWeyl}, for Szekeres Class I models the invariant scalar $\Psi_2$ is proportional to the density fluctuation, hence the gravitational thermodynamic variables become 
\begin{equation}
   \fl 8\pi \rho_{\textrm{\tiny{(gr)}}}=\frac{8\pi \alpha}{3}|\mathbf{D}_q^{\rho}|,\quad p_{\textrm{\tiny{(gr)}}} = 0, \quad q_a^{\textrm{\tiny{(gr)}}}=0, \quad 8\pi \Pi_{ab}^{\textrm{\tiny{(gr)}}} = \frac{4\pi \alpha}{3}|\mathbf{D}_q^{\rho}|(-2 \mathbf{r}_a\mathbf{r}_b+\mathbf{x}_a\mathbf{x}_b+\mathbf{y}_a\mathbf{y}_b).
\end{equation}
leading to the Gibbs-like equation 
\begin{equation}
    T_{\textrm{\tiny{(gr)}}}\dot{s}_{\textrm{\tiny{(gr)}}}= \partial_t\,(\rho_{\textrm{\tiny{(gr)}}} V) =\frac{\alpha}{3}\partial_t(|\mathbf{D}_q^{\rho}| a^3\mathcal{G}),
\end{equation}
while the gravitational temperature $T_{\textrm{\tiny{(gr)}}}$ is
\begin{equation}
    T_{\textrm{\tiny{(gr)}}}= \frac{|H+2\mathbf{D}_q^H|}{2\pi}=\frac{|H_q+3\mathbf{D}_q^H|}{2\pi},\label{gravT}
\end{equation}
resulting in the entropy production equation
\begin{equation}
    \dot{s}_{\textrm{\tiny{(gr)}}}=\frac{2\pi\alpha}{3}\frac{\partial_{t}(|\mathbf{D}_q^{\rho}| a^3\mathcal{G})}{|H_q+3\mathbf{D}_q^H|}.\label{dotsgrav}
\end{equation}
This enables us to analyze the conditions for positive entropy production.

\subsection{\label{sec:SdotPlus}Conditions for gravitational entropy production}

To arrive at the entropy production conditions for $\dot{s}_{\textrm{\tiny{(gr)}}} \geq 0$, which are analogous to the ones given by Sussman and Larena in \cite{Sussman:2013xpa,Sussman:2015bea}, we need the evolution equations for $a$, $\mathbf{D}^{\rho}_q$ and $\mathcal{G}$ given in equations \eref{eq:aev}, \eref{eq:Drhoev} and \eref{eq:Gev} respectively substitute them in the entropy production equation and after some algebraic manipulation,  we get three different cases for non-negative entropy production depending on the signs of $\mathbf{D}^{\rho}_q$ and $\mathbf{D}_q^H$
\begin{equation}
    \dot{s}_{\mbox{\tiny{(gr)}}} = \left\{ 
  \begin{array}{ c l }
    -\frac{2\pi \alpha a^3\mathcal{G}}{|H_q+3\mathbf{D}_q^H|}\rho_q\mathbf{D}_q^H & \quad \textrm{if } \mathbf{D}^{\rho}_q > 0 \\
    0                 & \quad \textrm{if } \mathbf{D}^{\rho}_q=0\\
    \frac{2\pi \alpha a^3\mathcal{G}}{|H_q+3\mathbf{D}_q^H|}\rho_q\mathbf{D}_q^H &\quad \textrm{if } \mathbf{D}^{\rho}_q < 0
  \end{array}
\right. ,
\end{equation}
this is because $a$, $\mathcal{G}$, $\alpha$, $\rho_q$, and the gravitational temperature $T_{\textrm{\tiny{(gr)}}}$ are already positive quantities. We can thus summarize the conditions for $\dot{s}_{\textrm{\tiny{(gr)}}}\geq 0$ as follows
\begin{equation}
\fl \dot{s}_{\mbox{\tiny{(gr)}}} = \left\{ 
  \begin{array}{ c l }
    \mathbf{D}_q^H < 0 & \quad \textrm{and } \mathbf{D}^{\rho}_q \geq 0 \\
    \mathbf{D}_q^H > 0 & \quad \textrm{and } \mathbf{D}^{\rho}_q \leq 0\\
  \end{array}
\right.
\Longrightarrow \dot{s}_{\textrm{\tiny{(gr)}}} \geq 0 \quad \mbox{if} \quad \mathbf{D}_q^H\mathbf{D}^{\rho}_q \leq 0.\label{eq:sdotplus}
\end{equation}
These conditions will be explored in the numerical example of \sref{section:numerical} in which we analyze a configuration of overdense structures around a void. 

\section{\label{sec:NumEx}Numerical example with two overdense structures around a central void}
\label{section:numerical}

In order to provide illustrative numerical examples of the CET gravitational entropy growth for arrays of jointly evolving structures, we introduce the following  dimensionless variables by normalizing with the Hubble parameter $\bar H_i$ of the FLRW background at the initial time slice $t=t_i$
\begin{eqnarray} 
\mathcal{H}_q &=& \frac{H_q}{\bar{H}_i},\qquad \tau = \bar{H}_i\,t,\qquad 
\Omega_q^m = \frac{8\pi \rho_q}{3\bar H_i^2},\qquad \Omega_q^k=\frac{K_q}{\bar H_i^2},\label{adimvarsdef}\\
{\bf D}^{m}_q &=& \frac{8\pi {\bf D}^{\rho}_q}{3\bar H_i^2},\quad {\bf D}^{k}_q = \frac{{\bf D}_q^{K}}{\bar H_i^2}, \quad {\bf D}^{h}_q = \frac{{\bf D}_q^{H}}{\bar H_i}.\label{adimfluctsdef}
 \end{eqnarray}
These dimensionless variables are convenient, not only because they facilitate the comparison of all Szekeres quantities with those of the background $\Lambda$CDM model (as they are normalized by $\bar H_i$), but also  because they allow for using the background redshift $z$ as a time parameter (from $\bar a=1/(1+z)$, where $\bar a$ is the background scale factor obtained in the asymptotic limit $a\to \bar a$ as $r\to\infty$ \cite{Sussman:2017otk}).
We also introduce for numerical convenience the ``stereographic'' spherical coordinates $[\tilde r,\theta,\phi]$  \cite{Plebanski:2006sd} related to the coordinates of \eref{eq:FLRWlike} by the transformation 
\begin{equation}
    \fl x = P(r)+S(r)\cot{(\theta/2)}\cos{(\phi)},\quad  y = Q(r)+S(r)\cot{(\theta/2)}\sin{(\phi)},\quad r = \tilde r,\label{eq:stereogr}
\end{equation}
where we will remove the tilde over $r$ to simplify notation. The transformation \eref{eq:stereogr} leads to the following non-diagonal metric that we have used in previous work on the models \cite{Sussman:2015fwa,Sussman:2015wna}
\begin{equation} ds^2 = -dt^2+a^2\,\gamma_{ij} dx^i dx^j,\qquad i,j=r,\,\theta,\,\phi \end{equation}
with $a= a(t,r)$ such that $a(t_i,r)=1$ for an initial time slice $t=t_i$ and 
\begin{eqnarray}
    \gamma_{rr} = \frac{(\Gamma-\mathbf{W})^2}{1-\bar H_i^2\Omega^k_{qi}r^2}+(\mathcal{N}+\mathbf{W}_{,\theta})^2+(1-\cos\theta)^2\,\mathcal{N}^2_{,\phi},\label{eq:stero1}\\
    \gamma_{r \theta}= -r(\mathcal{N}+\mathbf{W}_{,\theta}), \qquad  \gamma_{r \phi}=r\sin\theta(1-\cos\theta)\mathcal{N}^2_{,\phi},\label{eq:stero2}\\
    \gamma_{\theta \theta} = r^2, \qquad \gamma_{\phi \phi}=r^2\sin^2{\theta},\label{eq:stero3}
\end{eqnarray}
where the functions $\Gamma,\,\mathcal{N}$, the Szekeres dipole $\mathbf{W}$ and its magnitude $\mathcal{W}$ are are given by
\begin{eqnarray}
    \Gamma = 1+\frac{ra'}{r},\quad \mathcal{N}=\mathcal{X}\cos\phi+\mathcal{Y}\sin{\phi},\quad \mathbf{W}=-\mathcal{N}\sin{\theta}-\mathcal{Z}\cos{\theta}\label{eq:auxfuns}\\
 \quad \Longrightarrow \quad  \mathcal{W} = \frac{\pi}{2}\left[\int_0^\pi{\int_0^{2\pi}{\mathbf{W}^2 \rmd \theta \rmd \phi}}\right]^{1/2}=\sqrt{\mathcal{X}^2+\mathcal{Y}^2+\mathcal{Z}^2},\label{eq:dipolo}
\end{eqnarray}
with $\mathcal{X}=\mathcal{X}(r)$, $\mathcal{Y} = \mathcal{Y}(r)$, $\mathcal{Z}=\mathcal{Z}(r)$ arbitrary functions controlling the dipole direction, related to the functions $S,\,P,\,Q$ in \eref{eq:GammaEdefs} by
\begin{equation} 
\mathcal{X} = \frac{rP'}{S},\quad \mathcal{Y} = \frac{rQ'}{S},\quad \mathcal{Z} = \frac{rS'}{S},\quad
\end{equation}
so that setting $\mathcal{X}=\mathcal{Y}=\mathcal{Z}=0 \ \Longrightarrow \ \mathbf{W} = 0$ and results in a generic LTB seed model as the unique spherically symmetric subcase. This coordinate representation is very convenient to construct configurations consisting of multiple structures in a ``quasi-spherical'' arrangement \cite{Sussman:2015fwa,Sussman:2015wna}. 

\subsection{\label{sec:CondIn}Initial conditions}

We consider a set of two overdense structures (represented by a positive density contrast $\delta$) around a central void (represented by a negative density contrast $\delta$) inside a flat $\Lambda$CDM background. The background quantities at $z=0$, taken from \cite{Aghanim:2018eyx}, are 
\begin{equation}
    \fl \bar\Omega^{\textrm{\tiny{CDM}}}_0 = 0.3153, \quad \bar\Omega^{\Lambda}_0= 1- \bar\Omega^{\textrm{\tiny{CDM}}}_0 = 0.6847 \quad \mbox{and} \quad \bar H_0 = 67.36\, \frac{\hbox{km}}{\hbox{s\,Mpc}}.\label{obsparst0}
\end{equation}
For the initial time of evolution $t_i$ we choose the last scattering time at redshift  $z=1100$, to avoid effects of radiation and the coupled baryons. The subindex ${}_i$ will denote henceforth this initial time slice. Using the Friedman equation we obtain the background quantities for the initial time slice
\begin{equation}
    \fl \bar\Omega^{\textrm{\tiny{CDM}}}_i =  0.9999, \quad \bar\Omega^{\Lambda}_i= 1- \bar\Omega^{\textrm{\tiny{CDM}}}_0 \simeq 0 \quad \mbox{and} \quad \bar H_i = 1381799.54\, \frac{\hbox{km}}{\hbox{s\,Mpc}}.\label{obsparsLSS}
\end{equation}
The system \eref{eq:rhoev}-\eref{eq:DHconstr} remains identically valid for the dimensionless variables, with $\partial_\tau$ replacing the ``dot'' $\partial_t$ and with $a_i = \Gamma_i = \tilde{\mathcal{G}}_i=1$. The initial conditions are: values for the density, spatial curvature and Hubble q-scalars\footnote{In previous articles on LTB and Szekeres I models (for example \cite{Sussman:2013xpa,Sussman:2015bea,Sussman:2011bp}) we used a different notation for these initial value functions: $2\mu_{qi},\,\kappa_{qi}$ for $\Omega^m_{qi},\,\Omega^k_{qi}$, while the latter symbols were used to denote normalization with respect to $H_q^2$, as: $\Omega^m_{q}= 8\pi\rho_{q}/(3H_q^2),\,\Omega^k_{q}=K_q/H_q^2$ and $\Omega^\lambda = 8\pi\Lambda/(3H_q^2)$.}
\begin{equation}
    \Omega^m_{qi} = \frac{8 \pi \rho_{qi}}{3\bar H^2_i} \quad  \Omega^k_{qi} = \frac{K_{qi}}{\bar H^2_i},\quad \mathcal{H}_{qi} =\frac{H_{qi}}{\bar H_i}= \left[\Omega^m_{qi}-\Omega^k_{qi}+\bar\Omega_i^\Lambda\right]^{1/2},\label{adimvars}
\end{equation}
while the initial value fluctuations are
\begin{equation} \mathbf{D}^m_{qi}= \frac{r[\Omega^m_{qi}]'}{3(1-\mathbf{W})},\quad \mathbf{D}^k_{qi}= \frac{r[\Omega^k_{qi}]'}{3(1-\mathbf{W})},\quad \mathbf{D}_{qi}^h = \frac{\mathbf{D}_{qi}^m-\mathbf{D}_{qi}^k}{2\mathcal{H}_{qi}}.\label{adimflucts}
\end{equation}

\subsection{Setting up multiple structures}\label{multiple}

\begin{figure}[ht]
\centering
\begin{subfigure}{.5\textwidth}
  \centering
  \includegraphics[width=1\linewidth]{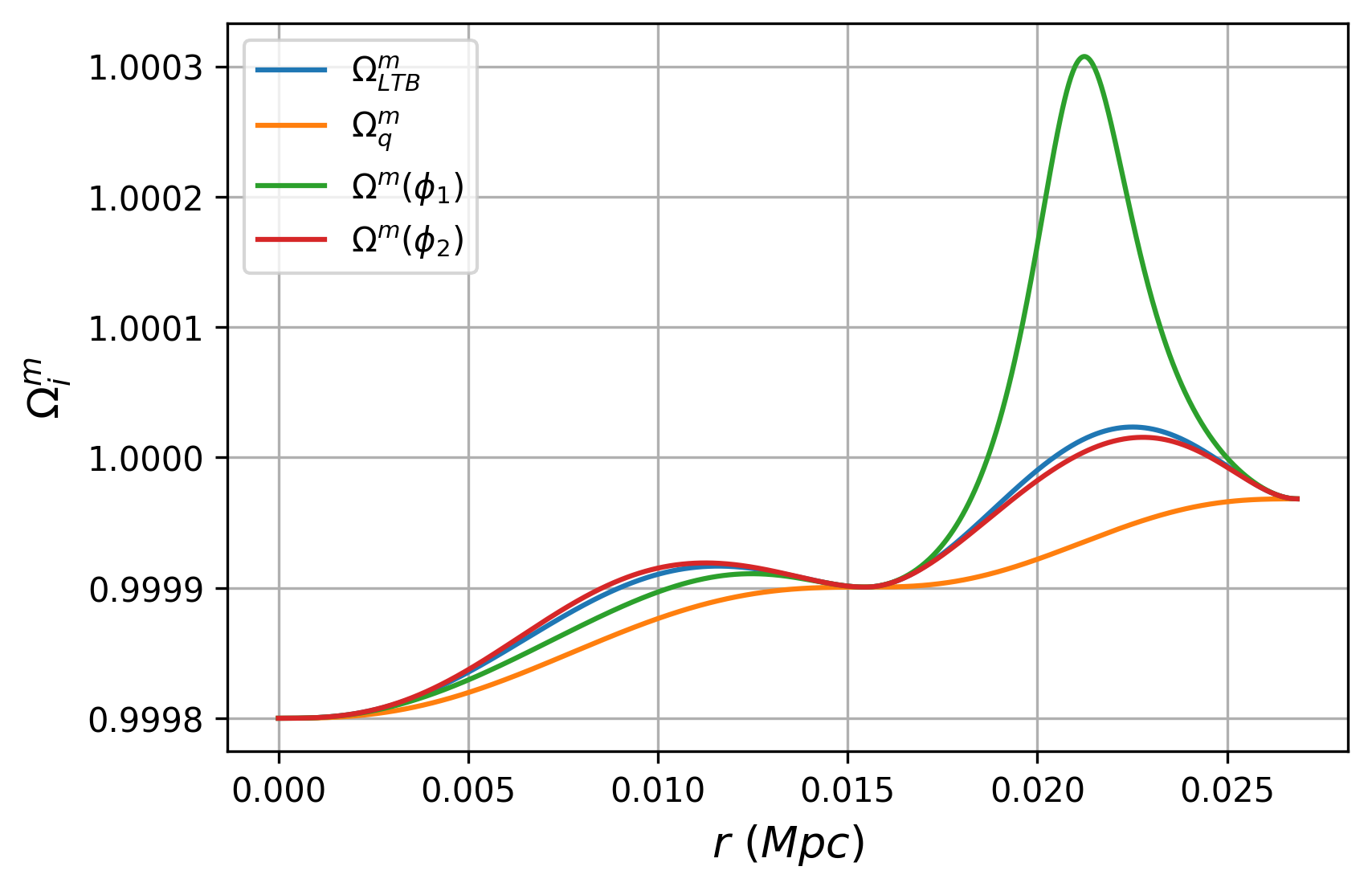}
  \caption{}
  \label{Fig:rhoi}
\end{subfigure}%
\begin{subfigure}{.5\textwidth}
  \centering
  \includegraphics[width=1\linewidth]{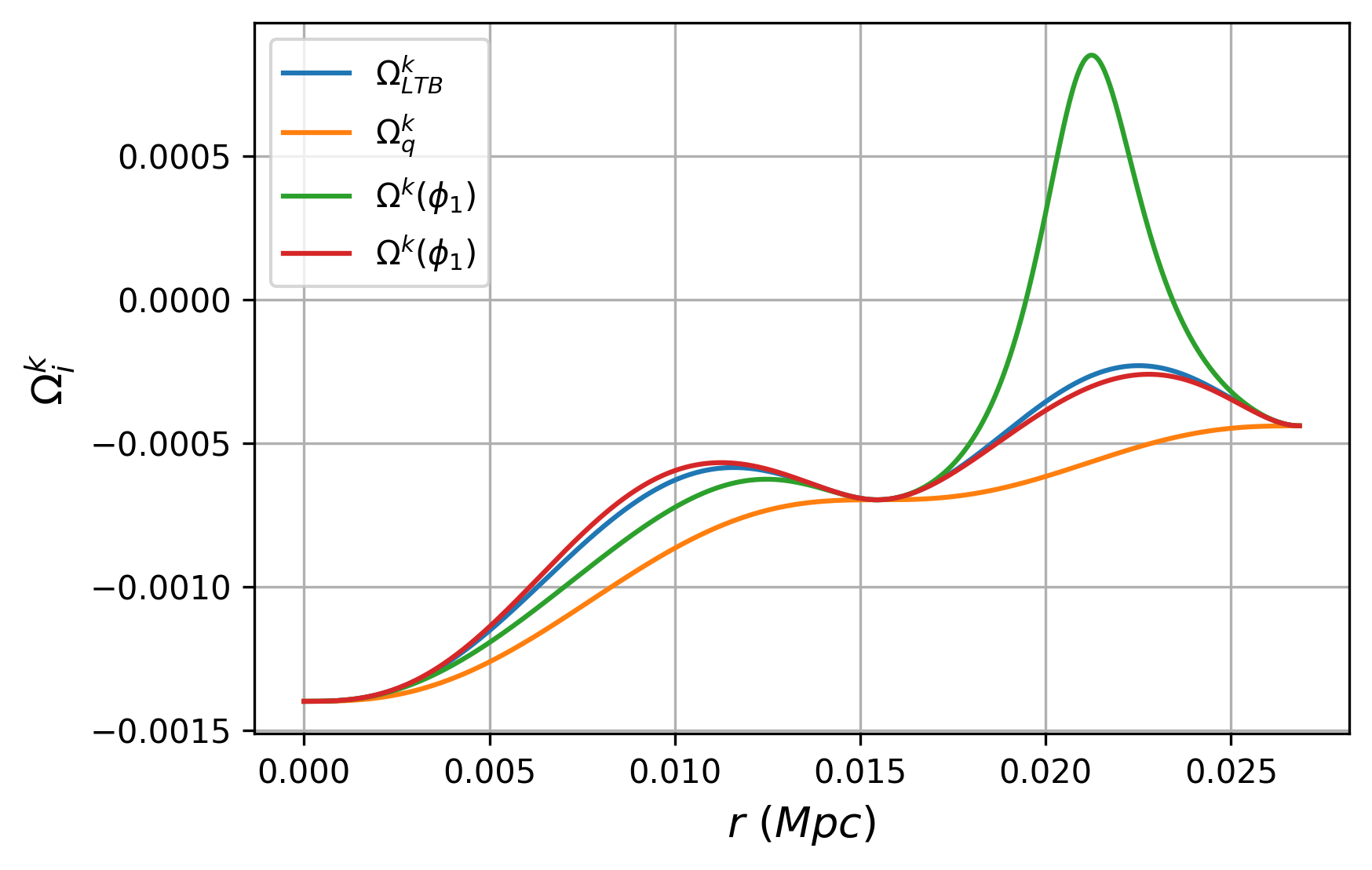}
  \caption{}
  \label{Fig:ki}
\end{subfigure}
\begin{subfigure}{.5\textwidth}
  \centering
  \includegraphics[width=1\linewidth]{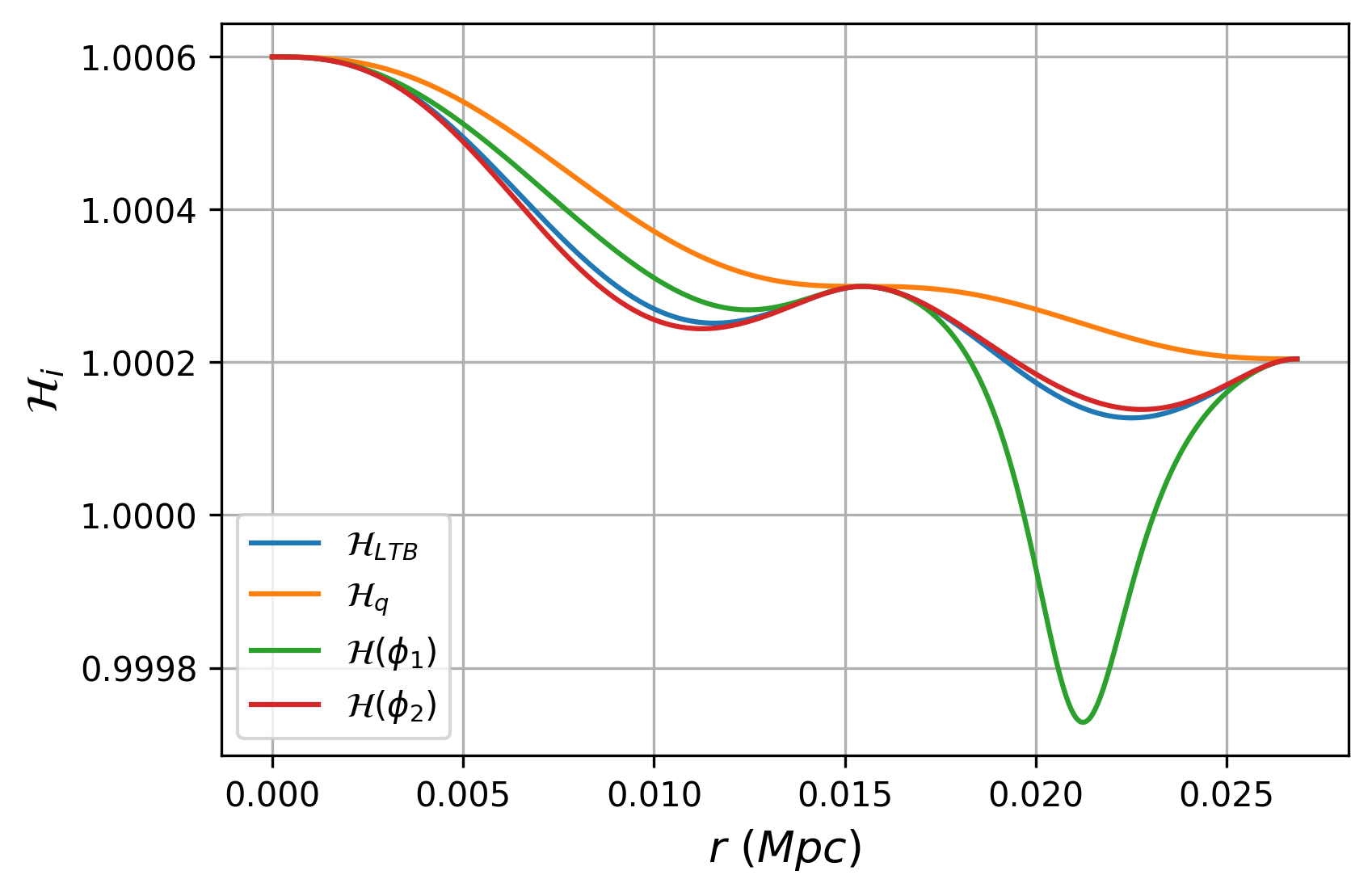}
  \caption{}
  \label{Fig:hi}
\end{subfigure}
\caption{Initial density $\Omega^{m}_{qi}$ (panel a), spatial curvature $\Omega^{k}_{qi}$ (panel b) and Hubble scalar $\mathcal{H}_{qi}$ (panel c) profiles computed from the expressions of \sref{multiple}. In all three panels the curves depict the profiles of the following quantities: the quasilocal scalars (yellow-orange curve) and the LTB scalars (blue curve) of the LTB seed model, while radial profiles of Szekeres I variables are depicted at two distinct radial rays with fixed angles (see figure \ref{Fig:Deltmi}) in the ``equator'' ($\theta=\pi/2$) of the coordinate system \eref{eq:stereogr}-\eref{eq:stero3}: $\phi_1$ from $r=0$ passing through the over-density (green curve) and $\phi_2$ from $r=0$ towards a region without visible structures (red curve). Note that these angles are distinct from the ones presented in \tref{table:1}. Notice how the green curves significantly deviate from the curves of the LTB seed model, while the red curves are almost indistinguishable from those of the LTB seed model. This shows how the angular dependence of the dipole $\mathbf{W}$ controls the deviation from spherical symmetry.}
\label{Fig:perfini}
\end{figure}

Without a stringent constraint of the degrees of freedom, we impose $\mathcal{Z}=0$. Next, we place a central void with initial density minimum at $r=0$, surrounded by two overdensities (density maxima) located at different radii and angles. We choose the two intervals such that $r= \chi r_s $ with $r_s = 0.0025$ Mpc, considering the intervals with boundaries given by
\begin{equation}
    \chi_0 = 0, \quad \chi_1 = 6.1823 \quad \mbox{and} \quad \chi_2 = 10.7376.
\end{equation}
In each of these intervals $\Omega_{qi}^m$ and $\Omega_{qi}^k$ are given as fifth order polynomials, $\mathcal{Q}_{i-1,i}$ and $\mathcal{P}_{i-1,i}$ respectively, that satisfy the following conditions which allow us to find the coefficients
\begin{eqnarray}
    \fl \mathcal{Q}_{i-1,i}(r_{i-1}) = 2m(r_{i-1}), \quad \mathcal{Q}_{i-1,i}(r_i) = 2m(r_{i}), \quad \mathcal{Q}'_{i-1,i} = \mathcal{Q}''_{i-1,i} = 0 \quad \mbox{at} \quad r=r_{i-1}, r_i,\nonumber\\
    &{}& \\
    \fl \mathcal{P}_{i-1,i}(r_{i-1}) = \kappa(r_{i-1}), \quad \mathcal{P}_{i-1,i}(r_i) = \kappa(r_{i}), \quad \mathcal{P}'_{i-1,i} = \mathcal{P}''_{i-1,i} = 0 \quad \mbox{at} \quad r=r_{i-1}, r_i,
\end{eqnarray}
where the auxiliary functions $m(r)$ and $\kappa(r)$ are defined by 
\begin{equation}
   m(r) = 0.5-\frac{0.0001}{1+[r/(r_1-r_0)]^3}, \quad  \kappa(r) =-\frac{0.0014}{1+[r/(r_1-r_0)]^{7/5}}.
\end{equation}
At radii larger than $r_2$ these functions converge into a $\Lambda$CDM background at $z=1100$ and, by themselves, represent a spherically symmetric void inside an FLRW background. In combination with the dipole and the fifth order polynomials, they define the location of the overdensities around the center of the void.  The dipole functions $\mathcal{X}$ and $\mathcal{Y}$ are chosen for each interval (these define the initial density maxima) as shown in \tref{table:1}.
\begin{table}
\centering
\begin{tabular}{ |c|c| }
 \hline
 \multicolumn{2}{|c|}{$r_0 < r < r_1 $} \\
 \hline
 $\mathcal{X}$ & $\mathcal{Y}$ \\
 \hline
 $-0.85f_1\cos{\left(\frac{5 \pi}{4}\right)}$   & $-0.85f_1\sin{\left(\frac{5 \pi}{4}\right)}$    \\
 \hline
 \multicolumn{2}{|c|}{$r_1 < r < r_2 $} \\
 \hline
 $\mathcal{X}$ & $\mathcal{Y}$ \\
 \hline
 $-0.8f_2\cos{\left(\frac{ \pi}{4}\right)}$   & $-0.8f_2\sin{\left(\frac{ \pi}{4}\right)}$\\
 \hline
\end{tabular}
\caption{Initial values for the dipole functions $\mathcal{X}$ and $\mathcal{Y}$ used in eqs. \eref{eq:auxfuns} and \eref{eq:dipolo} which define the angular location of the density maxima at each of the two radii intervals defined previously.}
\label{table:1}
\end{table}
The functions $f_i$ with $i=1,2$ in \tref{table:1} are defined as
\begin{equation}
    f_i(r) = \sin^2{\left[ \frac{(r-r_{i-1})\pi}{r_i-r_{i-1}} \right]},
\end{equation}
and rest of the variables are obtained from the constraints of the system of equations and the definitions of the fluctuations. The initial profiles obtained are displayed in figure \ref{Fig:perfini}.
%
%
\begin{figure}[ht]
\centering
\begin{subfigure}{.5\textwidth}
  \centering
  \includegraphics[width=1\linewidth]{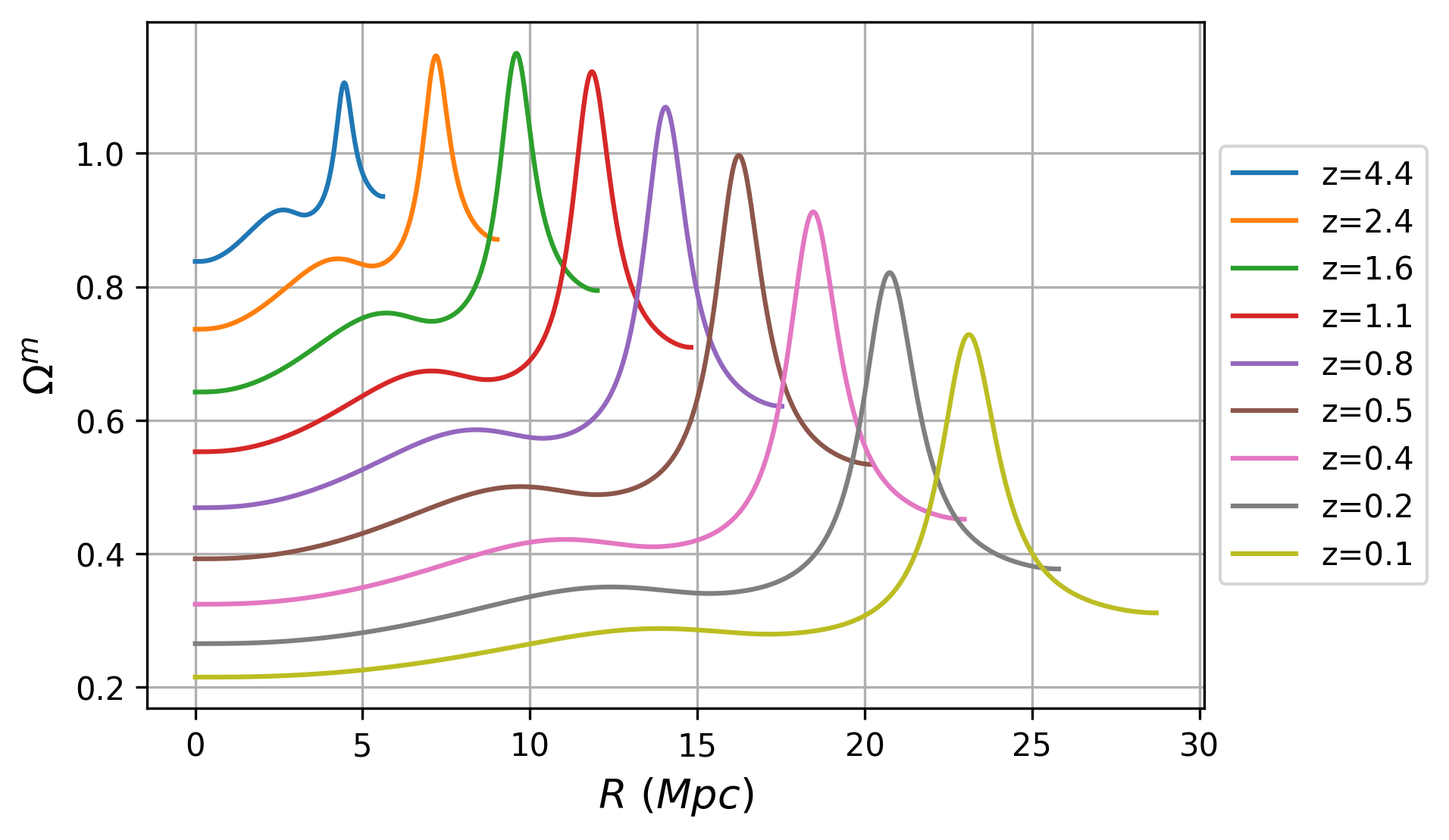}
  \caption{$\Omega^{m}(\phi_1)$.}
  \label{Fig:rhoqiRSD}
\end{subfigure}%
\begin{subfigure}{.5\textwidth}
  \centering
  \includegraphics[width=1\linewidth]{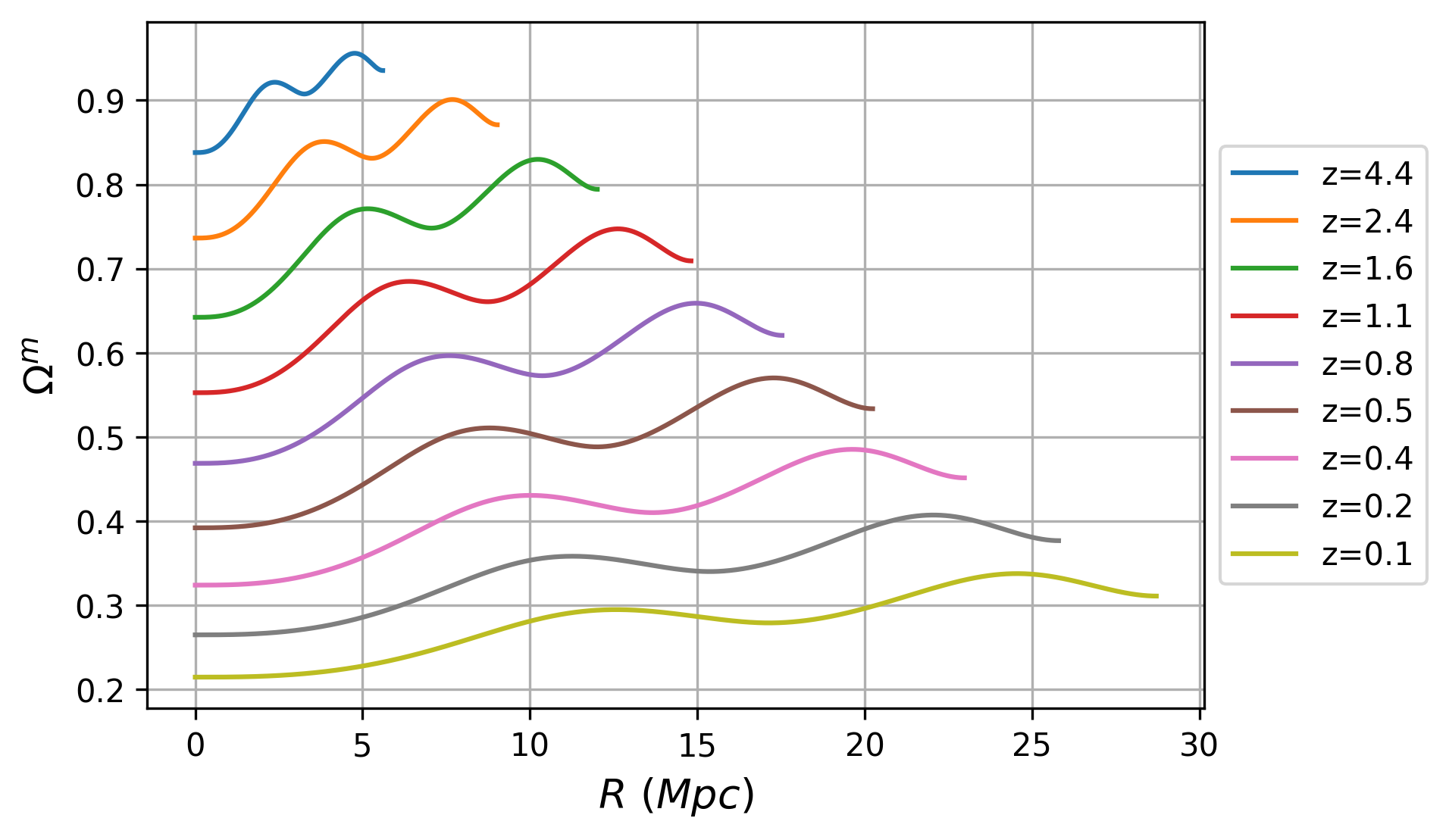}
  \caption{$\Omega^{m}(\phi_2)$.}
  \label{Fig:rhoqiRV}
\end{subfigure}
\begin{subfigure}{.5\textwidth}
  \centering
  \includegraphics[width=1\linewidth]{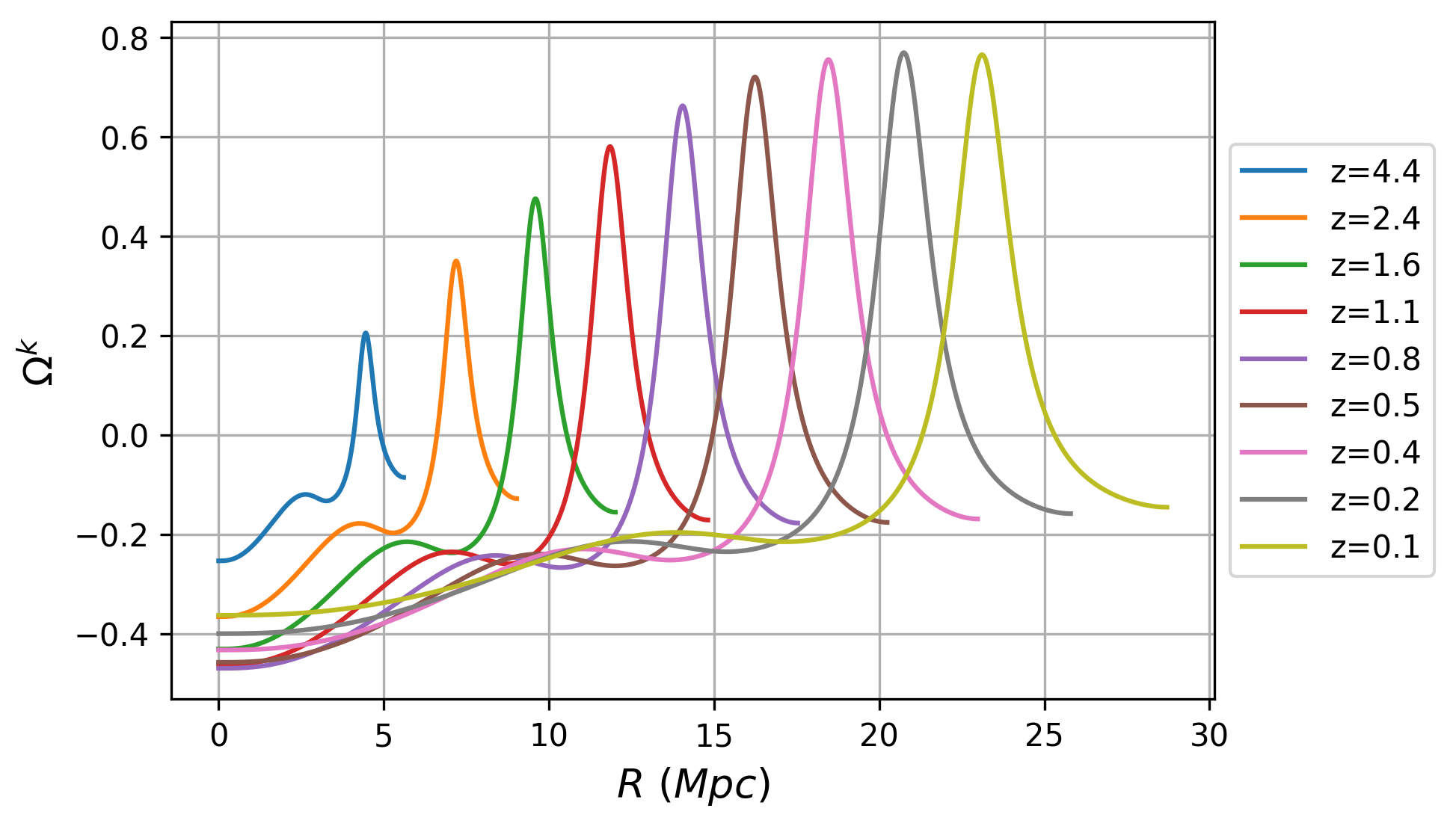}
  \caption{$\Omega^{k}(\phi_1)$.}
  \label{Fig:kiRSD}
\end{subfigure}%
\begin{subfigure}{.5\textwidth}
  \centering
  \includegraphics[width=1\linewidth]{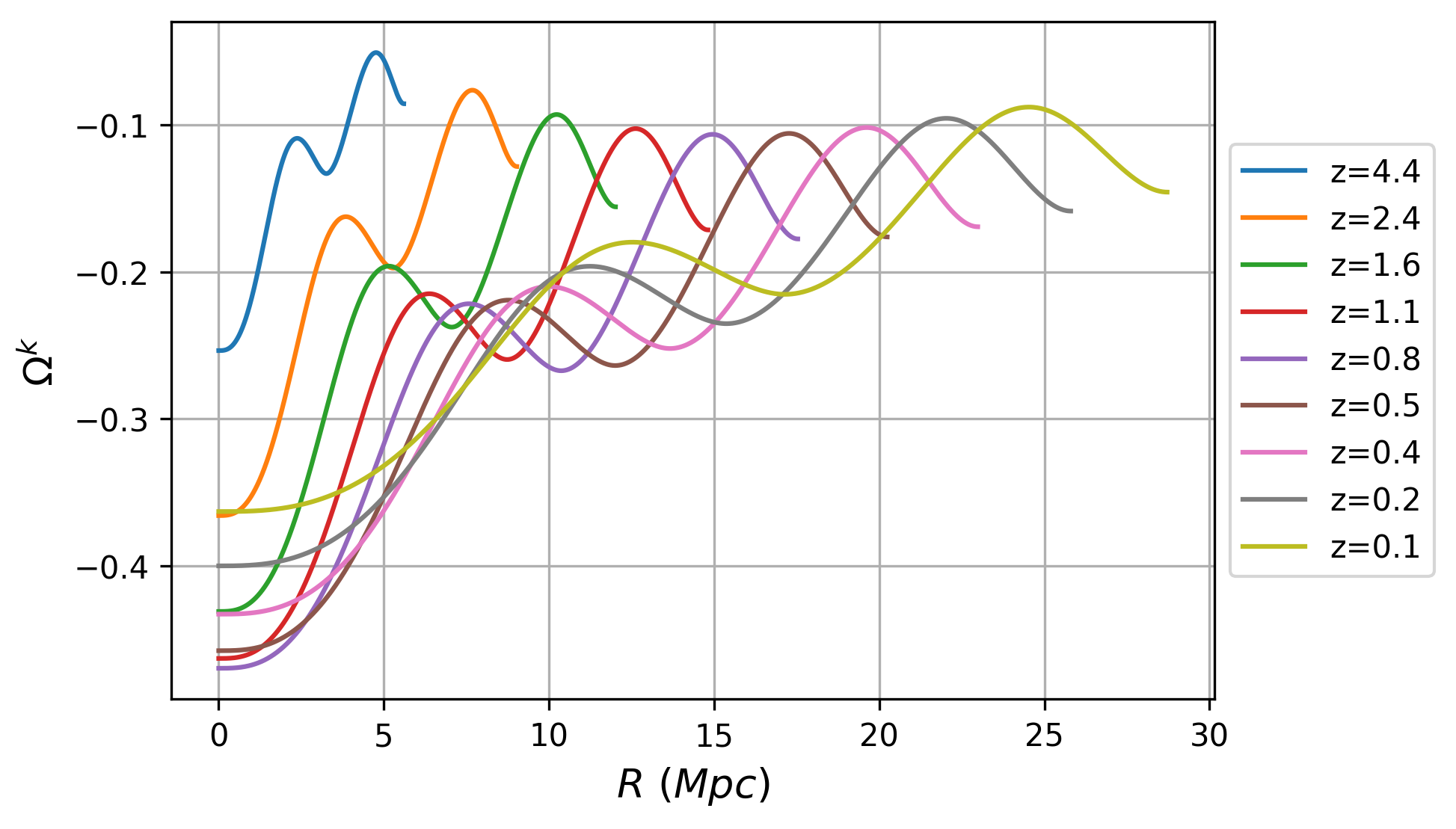}
  \caption{$\Omega^{k}(\phi_2)$.}
  \label{Fig:kiRV}
\end{subfigure}
\begin{subfigure}{.5\textwidth}
  \centering
  \includegraphics[width=1\linewidth]{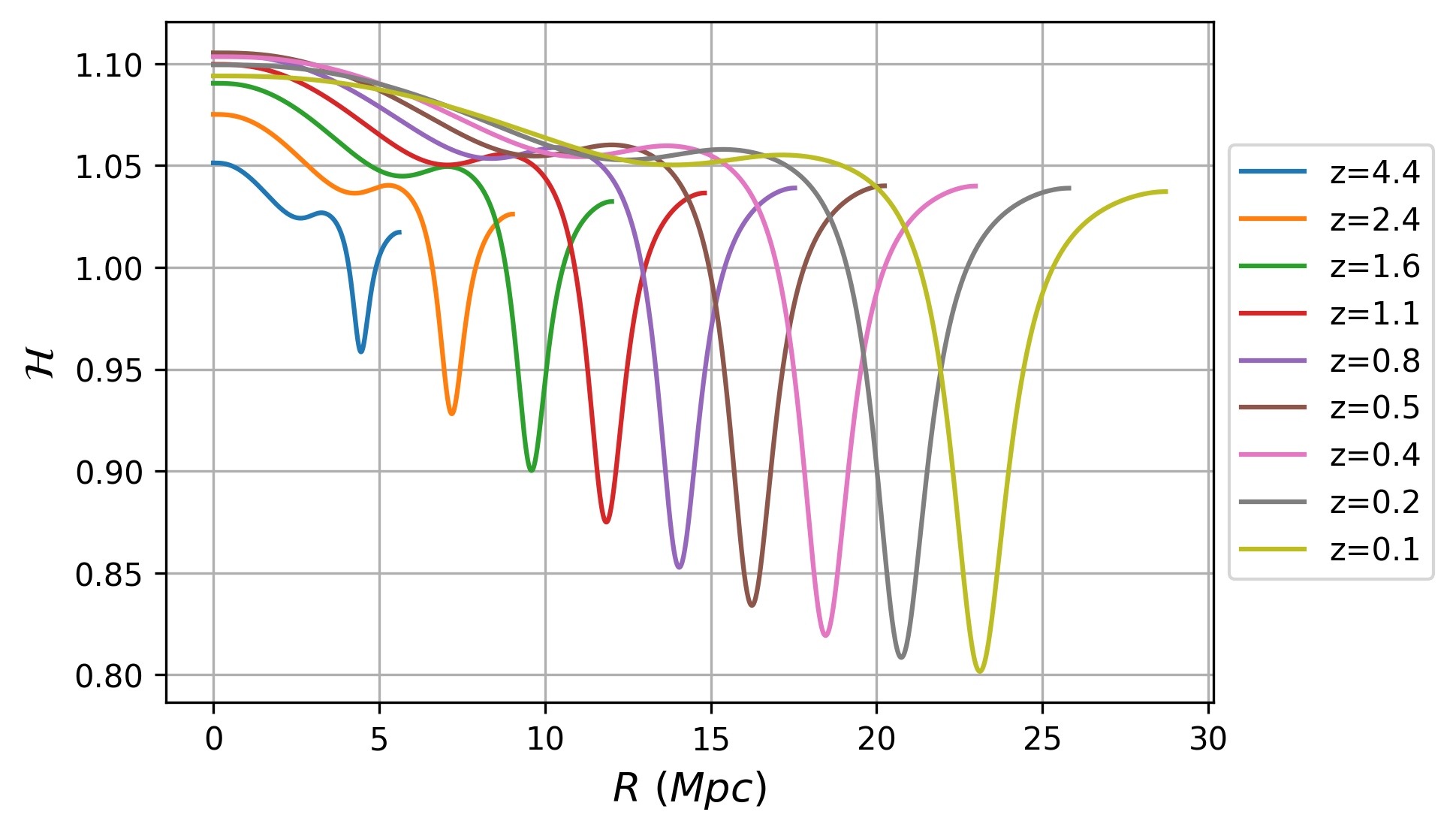}
  \caption{$\mathcal{H}(\phi_1)$.}
  \label{Fig:hiRSD}
\end{subfigure}%
\begin{subfigure}{.5\textwidth}
  \centering
  \includegraphics[width=1\linewidth]{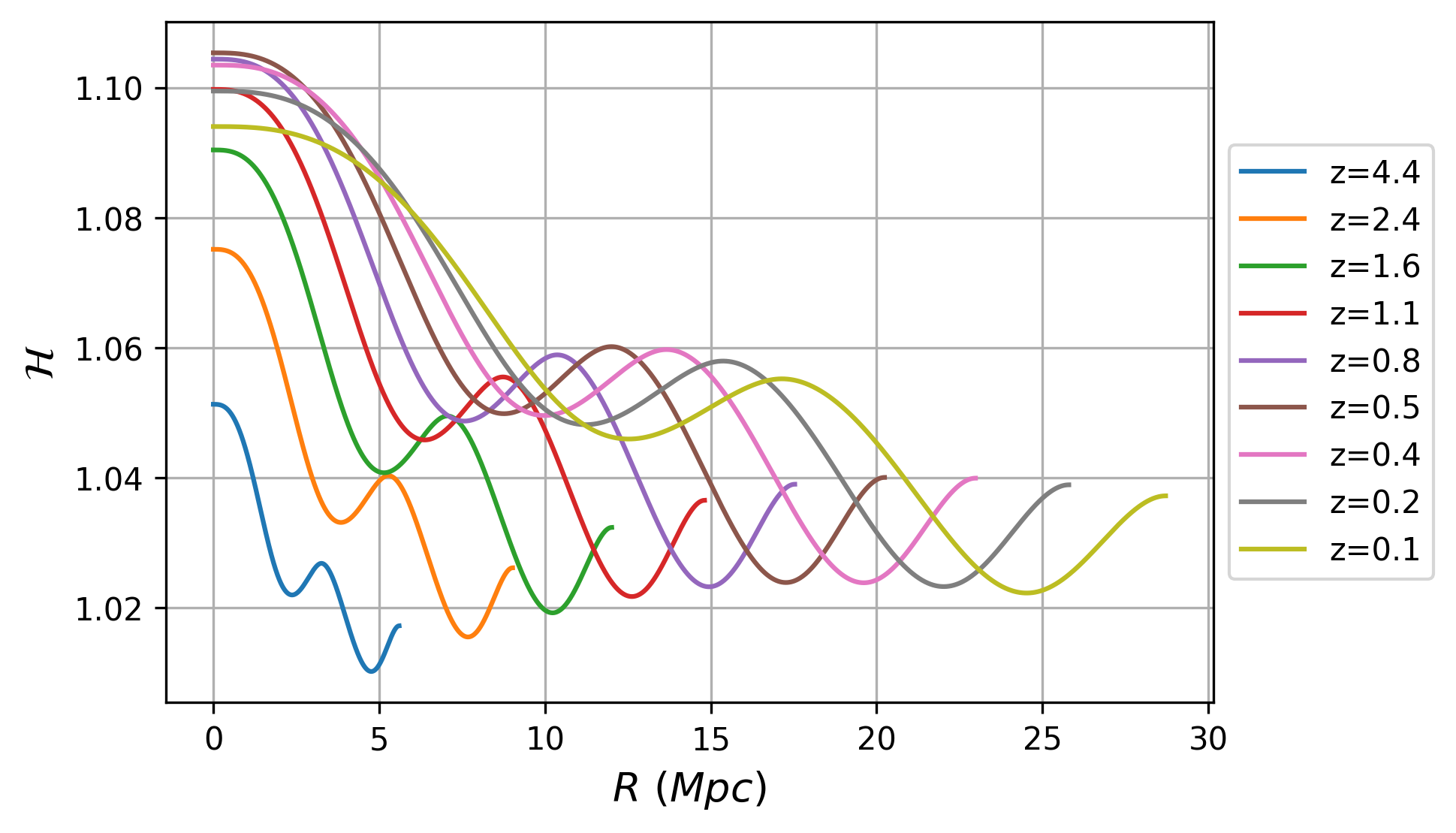}
  \caption{$\mathcal{H}(\phi_2)$.}
  \label{Fig:hiRV}
\end{subfigure}
\caption{Evolution of the density ($\Omega^m =8\pi\rho/(3\bar H^2)$ panels a, b), spatial curvature ($K/\bar H^2$ panels c, d) and Hubble scalar ($\mathcal{H}=H/\bar H$ panels e, f) in the two fixed angles $\phi_1$ (left hand side panels) and $\phi_2$ (right hand side panels) used in figure \ref{Fig:perfini} (see depiction of the angles in figure \ref{Fig:Deltasq}). The scalars are plotted as a function of $R(t,r)=r\,a(t,r)$ to show how the initial configurations grow in amplitude and size scale as the model evolves between $z=4.4$ and $z=0.1$. As reference, at the initial time slice $z=1100$ the location of the density maximum for the profile at $\phi_1$ is located at  $R_i=r=0.021\ Mpc$.}
\label{Fig:perfsev}
\end{figure}
\FloatBarrier

\subsection{\label{sec:NumRes}Numerical results}

The system of equations \eref{eq:rhoev}-\eref{eq:Gev} is solved using the finite difference method as described in \cite{alcubierre2008introduction}\footnote{Our code is written in the Python language. It's inspired in that of Ref. \cite{Sussman:2015wna} and is available in the  following URL: \url{https://github.com/klesto92/SzekeresIEntropy}.}. We can appreciate the time and scale evolution of the three main covariant scalars: density, spatial curvature and Hubble scalar ($\Omega_q^m,\,\Omega_q^k\,\mathcal{H}_q$) by plotting them in figure \ref{Fig:perfsev} as functions of $R=a(t,r)\,r$ for redshifts between $z=4.4$ to $z=0.1$ along two distinct radial rays with fixed angles in the ``equator'' ($\theta=\pi/2$) of the coordinate system \eref{eq:stereogr}-\eref{eq:stero3}, $\phi_1$ from $r=0$ passing through the over-density and $\phi_2$ from $r=0$ towards a region without visible structures, as in figure \ref{Fig:perfini} (see depiction of the angles in figure \ref{Fig:Deltmi}). As shown in figure \ref{Fig:rhoqiRSD}, an  over-density located at $R=5$ Mpc from the worldline $r=0$ at $z=4.4$ in the angular direction $\phi_1$ grows in amplitude and is located at $R=23$ Mpc at $z=0.1$. As a contrast, the density profile along $\phi_2$ (right hand side panels) barely grows from $z=4.4$ to $z=0.1$. Similar evolution patterns occur for the other scalars. In particular, figures \ref{Fig:hiRSD} and \ref{Fig:kiRSD} show how in the over-dense region the Hubble scalar significantly decreases, while spatial curvature increases and becomes positive. 

A more comprehensive way to examine the growth of structures follows by plotting variables in the ``equatorial'' plane defined by $\theta=\pi/2$ in the spherical coordinates \eref{eq:stereogr}-\eref{eq:stero3}.  While the figures we present in this section seem to be given as equatorial planes in flat polar coordinates, this is merely a coordinate effect, as in Szekeres I models the time slices (and in particular the ``equator'' $\theta=\pi/2$) have non-trivial curvature that deviates from spherical symmetry, as they are foliated by non-concentric 2-spheres (see extensive discussion in \cite{Sussman:2015fwa,Sussman:2015wna}).
\FloatBarrier
\begin{figure}[ht]
\centering
\begin{subfigure}{.5\textwidth}
  \centering
  \includegraphics[width=0.9\linewidth]{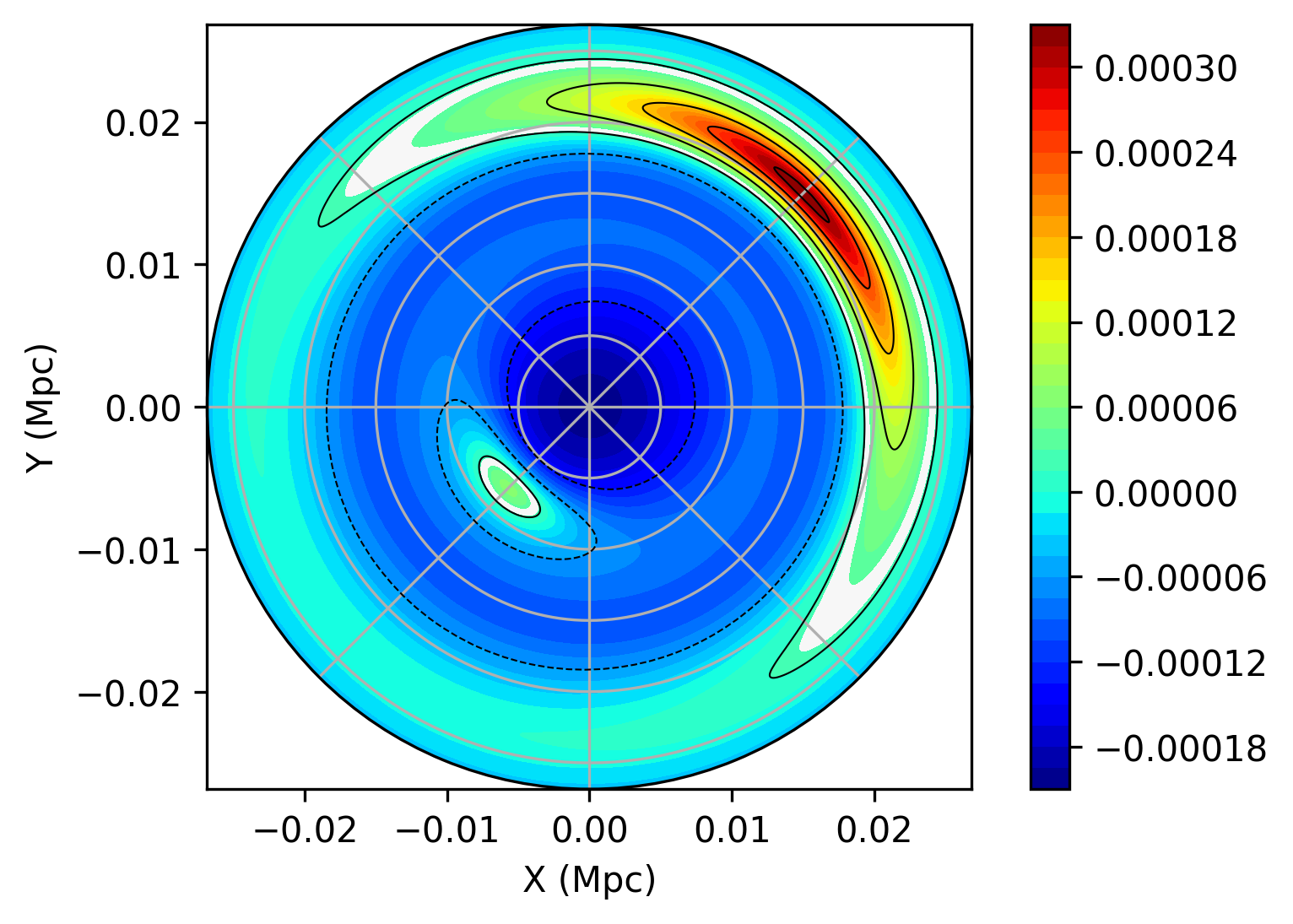}
  \label{Fig:deltai}
\end{subfigure}%
\begin{subfigure}{.5\textwidth}
  \centering
  \includegraphics[width=0.9\linewidth]{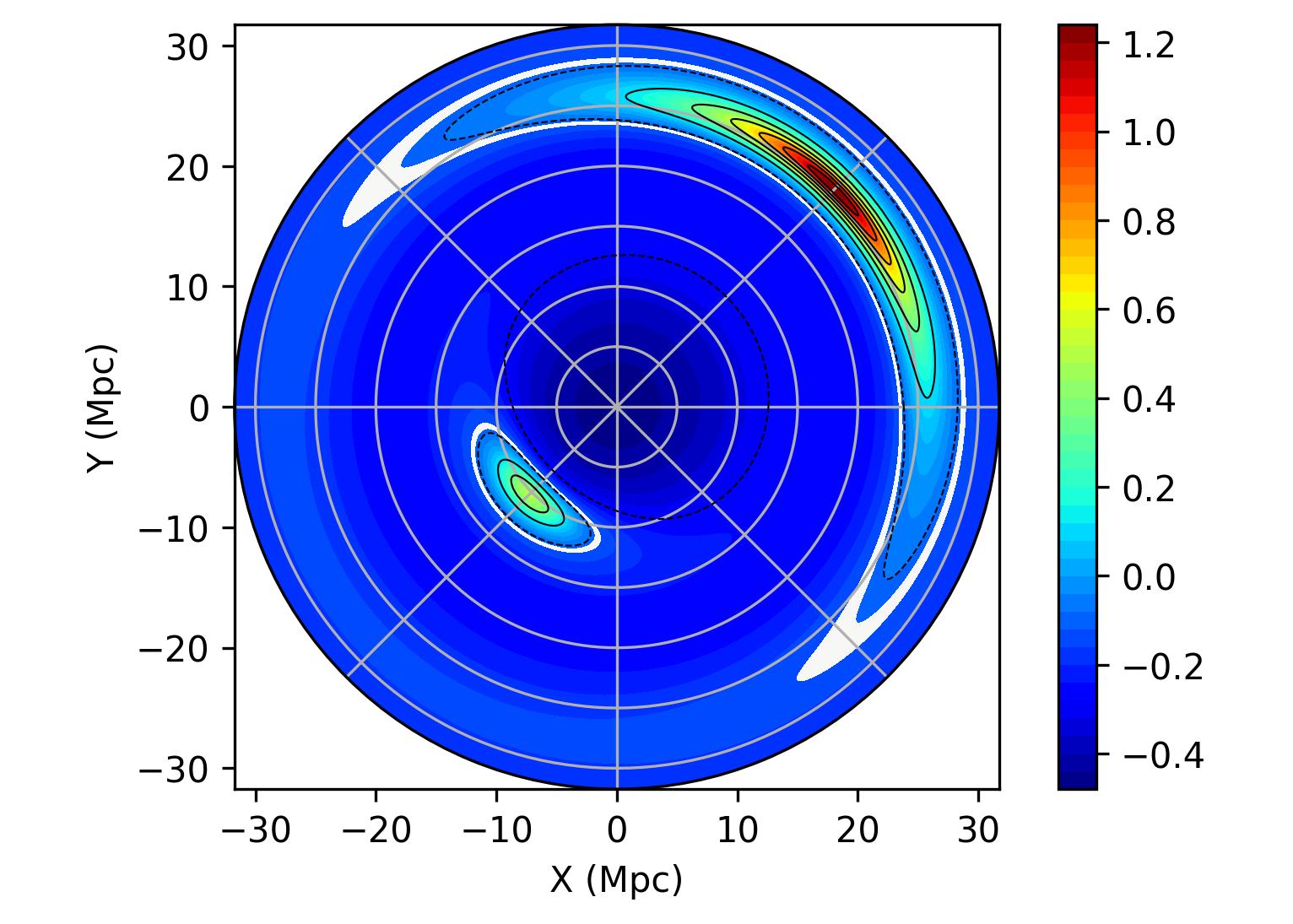}
  \label{Fig:deltaf}
\end{subfigure}
\caption{Initial and final density contrast $\delta$ of the configuration defined in section \ref{multiple}. The left panel corresponds to the initial configuration at $z=1100$ while the right one is the final configuration at $z=0$. For illustrative purposes we show the configuration in the equatorial plane. Notice how the region around the center reaches negative values of $\delta$ signaling the presence of a void, while $\delta$ around the maxima at the different angles achieves positive values, these regions delimited by the white bands are what we refer to as the over-dense structures.}
\label{Fig:deltas}
\end{figure}
%
\begin{figure}[ht]
\centering
\begin{subfigure}{.5\textwidth}
  \centering
  \includegraphics[width=0.9\linewidth]{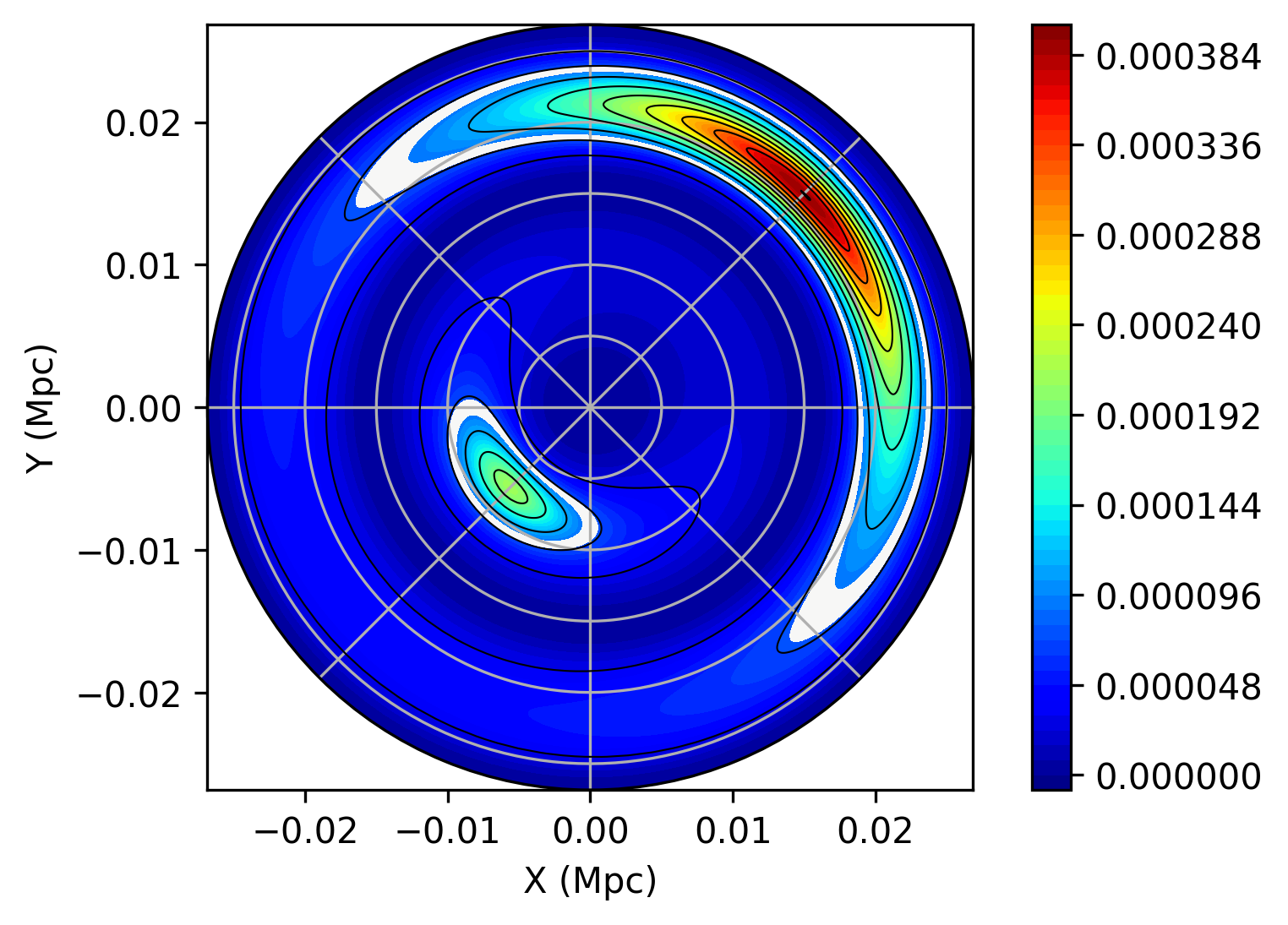}
  \caption{$\mathbf{D}^{m}_q$ at $z=1100$.}
  \label{Fig:Dmi}
\end{subfigure}%
\begin{subfigure}{.5\textwidth}
  \centering
  \includegraphics[width=0.9\linewidth]{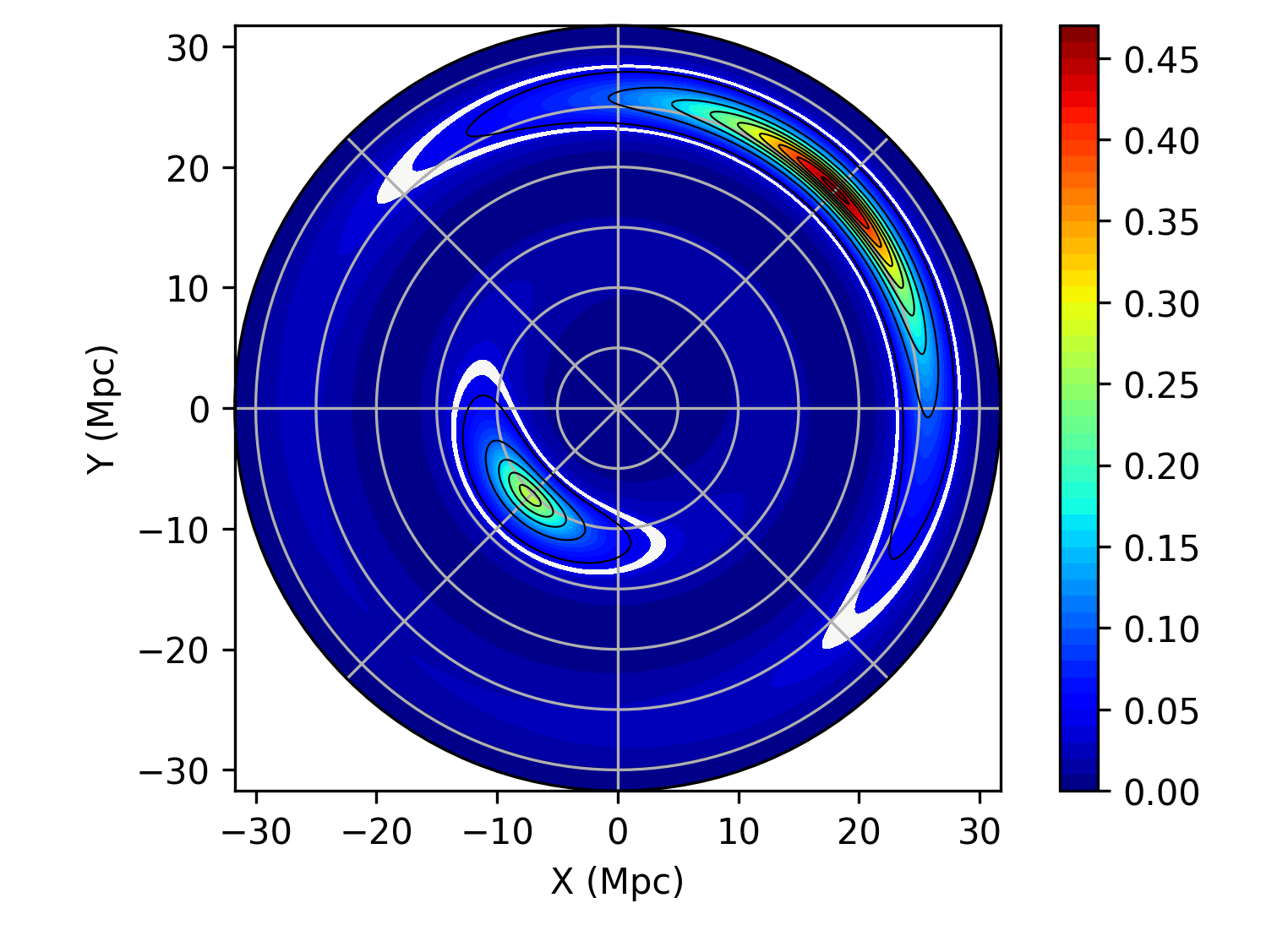}
  \caption{$\mathbf{D}^{m}_q$ at $z=0$.}
  \label{Fig:Dmf}
\end{subfigure}
\begin{subfigure}{.5\textwidth}
  \centering
  \includegraphics[width=0.9\linewidth]{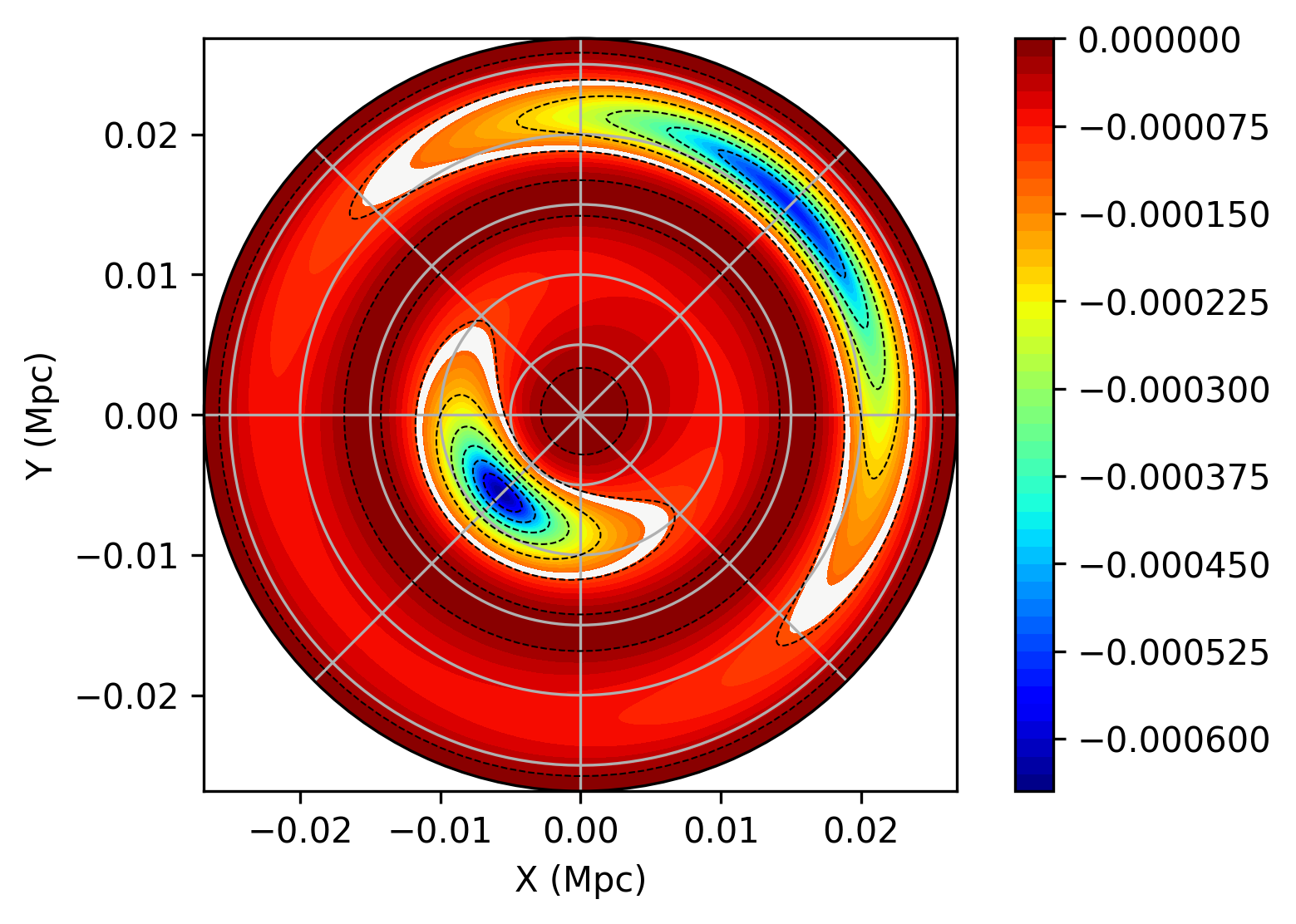}
  \caption{$\mathbf{D}^{h}_q$ at $z=1100$.}
  \label{Fig:Dhi}
\end{subfigure}%
\begin{subfigure}{.5\textwidth}
  \centering
  \includegraphics[width=0.9\linewidth]{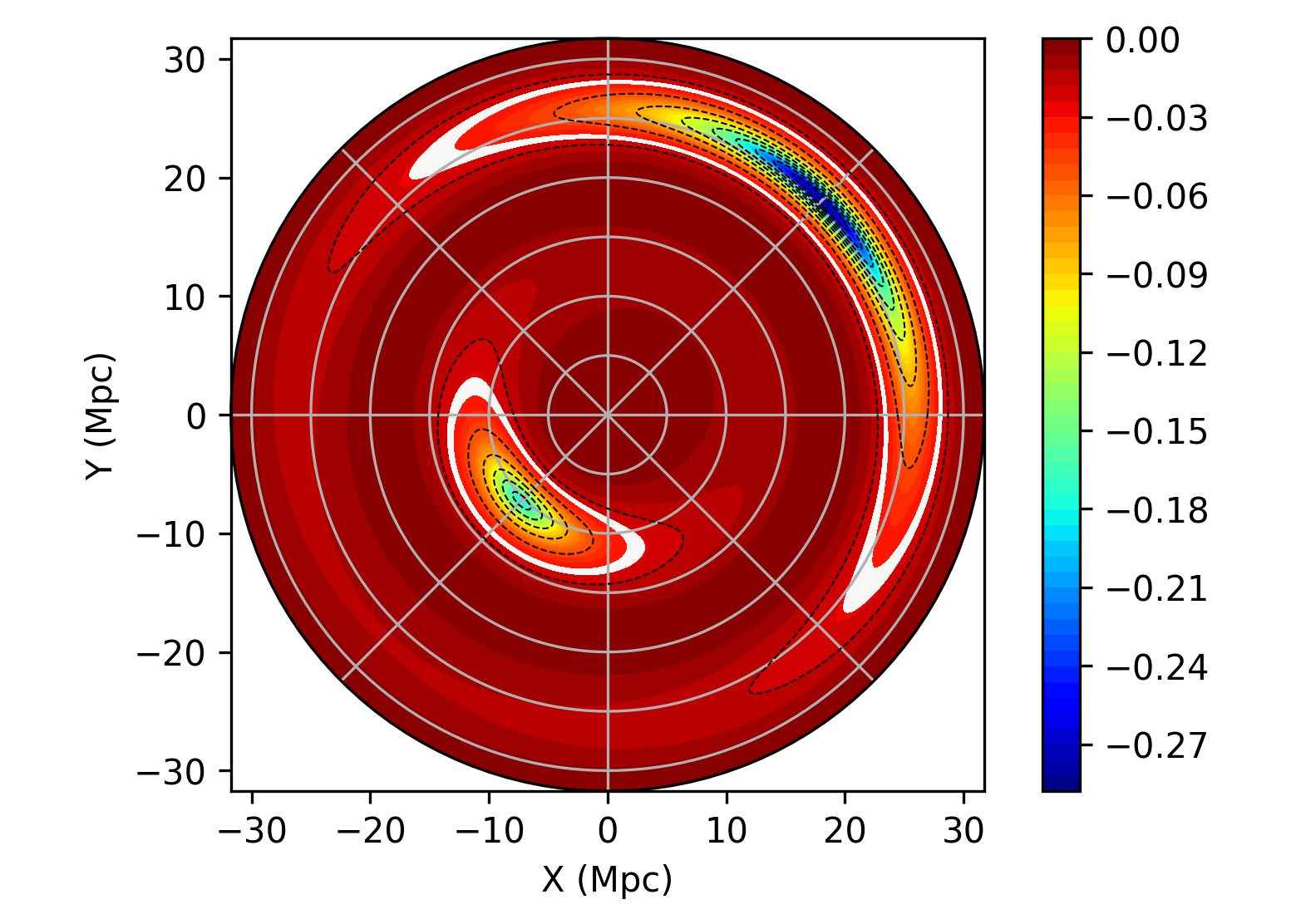}
  \caption{$\mathbf{D}^{h}_q$ at $z=0$.}
  \label{Fig:Dhf}
\end{subfigure}
\caption{As in figure \ref{Fig:deltas}, the left side panels correspond to the initial configuration and the right side panels to the final configuration. Panels a and b correspond to the density fluctuation $\mathbf{D}^{m}_q$, panels c and d correspond to the expansion fluctuation $\mathbf{D}^{h}_q$. The regions delimited by the white bands are the over-dense structures.}
\label{Fig:DmYDh}
\end{figure}
\begin{figure}[ht]
\centering
\begin{subfigure}{.5\textwidth}
  \centering
  \includegraphics[width=0.85\linewidth]{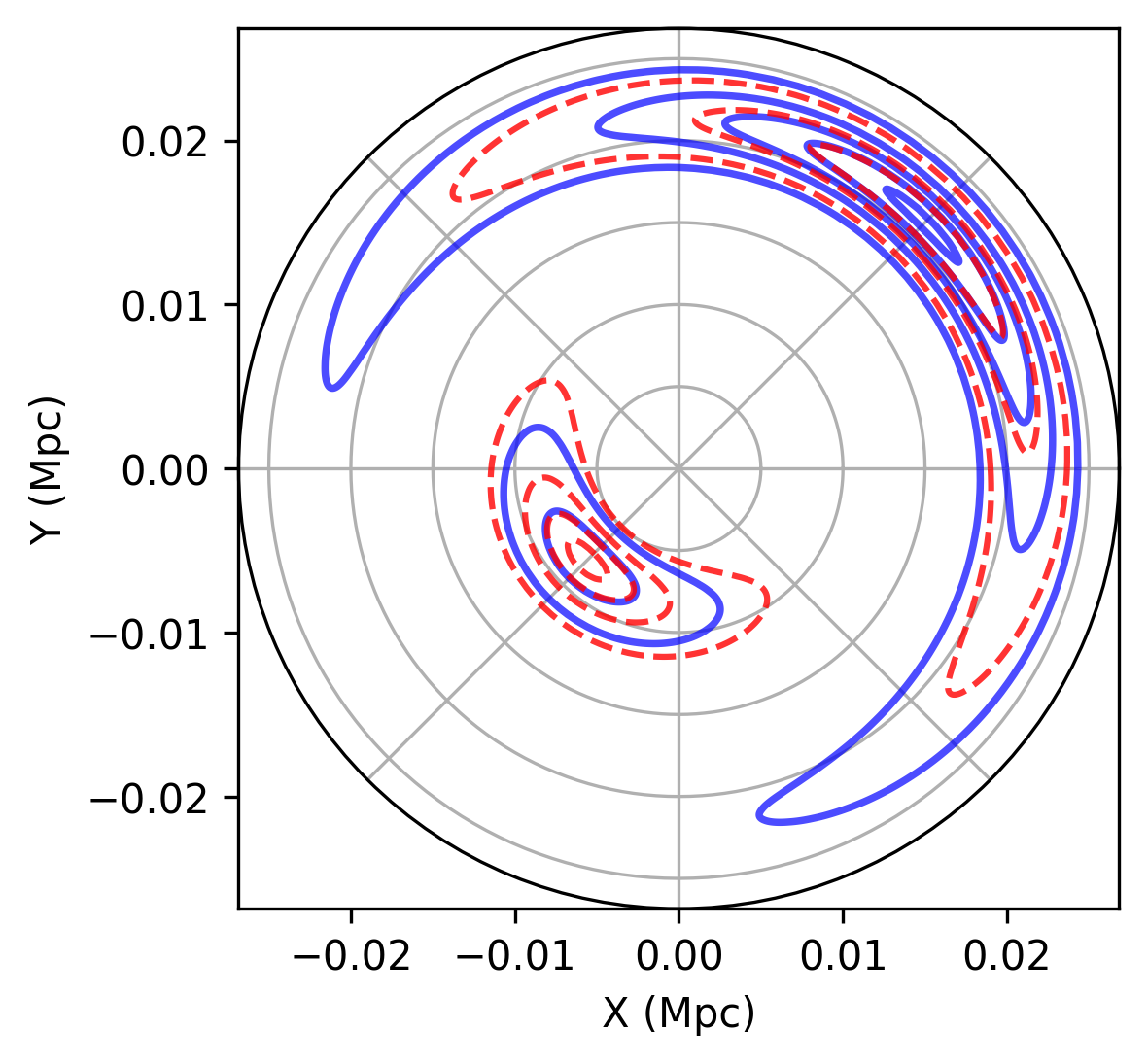}
  \label{Fig:DmDhi}
\end{subfigure}%
\begin{subfigure}{.5\textwidth}
  \centering
  \includegraphics[width=0.85\linewidth]{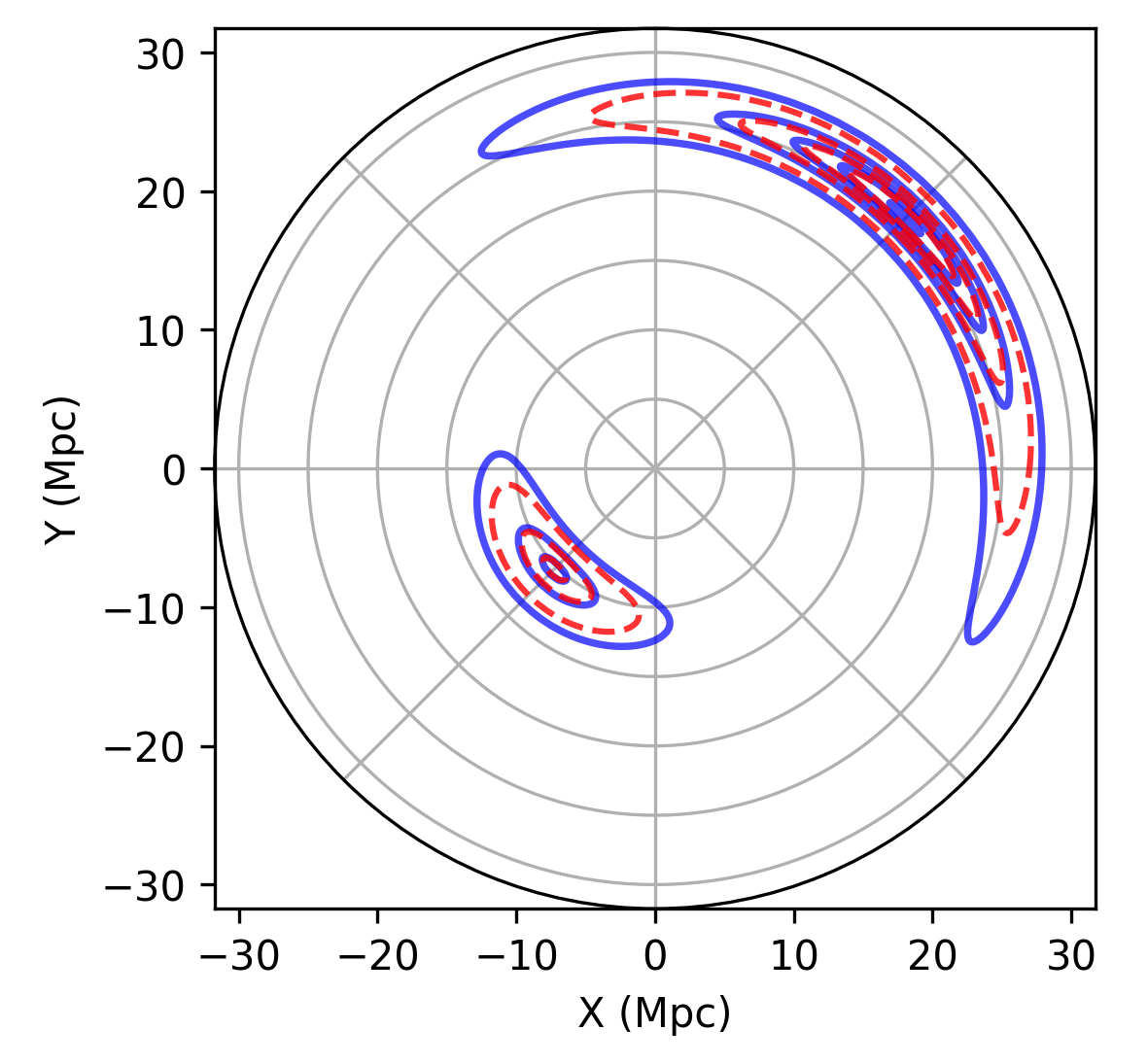}
  \label{Fig:DmDhf}
\end{subfigure}
\caption{Some of $\mathbf{D}^{m}_q$'s (blue) and $\mathbf{D}^{h}_q$'s (red) level curves superposed. As before, the left side panel is the initial configuration and the right side panel shows the final configuration. Each plot corresponds to the example in the figures \ref{Fig:deltas} and \ref{Fig:DmYDh}, which point at the regions where $\mathbf{D}^{m}_q$ and $\mathbf{D}^{h}_q$ have opposite signs, corresponding to positive entropy production as dictated by equation \eref{eq:sdotplus}.}
\label{Fig:deltascurvas}
\end{figure}
\FloatBarrier
The left hand side panel of figure \ref{Fig:deltas} displays the density contrast defined as $\delta = \rho/\bar{\rho}-1$, where $\bar{\rho}$ is the background matter density at the present cosmic time $z=0$. The system is evolved from the last scattering surface up to the present time and the result is shown in the right panel of the same figure. We can clearly distinguish the central void and the surrounding over-dense structures by their colors, with the stark red regions being the maximum values, and the dark blue regions being the minimum. Another property displayed throughout the evolution is that the whole configuration grows from a radius of $\sim 0.0025Mpc$ to a radius of $\sim 30Mpc$. 

Note that the configuration grows not only in size but in amplitude too, as seen in the color bar besides each panel in the figure. This will be the case for all the variables we present in this work. Now we turn our attention to the quantities relevant to the entropy production: $\mathbf{D}^{m}_q$ and $\mathbf{D}^{h}_q$.

The results for the numerical evolution of the $\mathbf{D}^{m}_q$ and $\mathbf{D}^{h}_q$  are given in figure \ref{Fig:DmYDh}. As mentioned in \cite{Sussman:2015fwa}, these figures satisfy at $r=0$ (where the center of the void lies) $\mathbf{D}^{m}_q=\mathbf{D}^{h}_q=0$. From here and into the overdense structures at the different angular and radial locations the fluctuations take either positive (for the density fluctuation) or negative (for the expansion fluctuation) values.
The starker red color in both the top and lower panels represent the highest values taken by each of the variables while the darker blue color represents the lowest values achieved. Therefore in the case of the $\mathbf{D}^{h}_q$, this particular example shows only negative values with the maximum being $0$, the structure's values go from a maximum represented by dark red to a minimum represented by dark blue. In the case of $\mathbf{D}^{h}_q$ its minimum is dark blue where this particular fluctuation takes negative values. For the density fluctuation $\mathbf{D}^{m}_q$ the opposite happens. This leads us to the desired result of a positive entropy production in equation \eref{eq:sdotplus}. We show such a result in figure \ref{Fig:deltascurvas}  where we superimposed the structures' level curves of both fluctuations on top of each other: these are the regions where both fluctuations have opposite signs. As a result we can conclude that the entropy production is positive for all Szekeres I models.

\section{\label{sec:Disc}Discussion of Results}

 While the figures of \sref{sec:NumRes} show convincingly that the regions satisfying entropy growth conditions \eref{eq:sdotplus} coincide with the regions where structures are located, we have not considered some important theoretical issues, such as:  (i) the asymptotic behavior, not only of $\dot{s}_{\textrm{\tiny{(gr)}}}$ but of the gravitational entropy $s_{\textrm{\tiny{(gr)}}}$ itself and (ii) the relation between entropy growth and the density growing and decaying modes. Since these issues were examined previously for generic LTB models \cite{Sussman:2013xpa} and for spherical LTB voids \cite{Sussman:2015bea}, we can simplify the discussion by looking at how the Szekeres I variables relate to their equivalents in their LTB seed model. 
 
Quantities such as $a,\,\Gamma$ and the q-scalars $\Omega_q^m,\,\Omega_q^k,\,\mathcal{H}_q$ are independent of the dipole parameters, depending only on $(t,r)$, thus are common to their equivalents in the LTB seed model obtained by setting $\mathbf{W}=0$. However, the Szekeres I fluctuations $\mathbf{D}^{m}_q$ and $\mathbf{D}^{h}_q$ explicitly depend on $\mathbf{W}$, their relation with their LTB counterparts that follow from $\mathbf{W}=0$ is given by 
\begin{eqnarray} 
\mathbf{D}^{A}_q &=& \frac{\Gamma}{\Gamma-\mathbf{W}}\,\mathbf{d}_q^{A},\qquad \Delta_q^A = \frac{\Gamma}{\Gamma-\mathbf{W}}\,\delta_q^A\nonumber\\
 \hbox{with} &{}&\quad \mathbf{d}_q^{A}=\mathbf{D}_q^{A}|_{\mathbf{W}=0},\qquad \delta_q^A=\Delta_q^{A}|_{\mathbf{W}=0}\quad \hbox{for}\quad A = m,\,k,\,h.\label{Ddrelation}
\end{eqnarray}  
We are assuming absence of shell crossings by demanding that  $\Gamma>0$ and $\Gamma-\mathbf{W}>0$ must hold everywhere, so that $\mathbf{D}^{A}_q$ and $\mathbf{d}^{A}_q$ are fully regular and always have the same sign and qualitatively similar evolution (a shell crossing can be understood as marking the breaking down of the validity of dust as a matter-energy model for CDM). As a consequence, Szekeres I models directly inherit the gravitational entropy conditions of their LTB seed model: $\mathbf{D}^{m}_q\,\mathbf{D}^{h}_q\leq 0\,\,\Leftrightarrow \,\,\mathbf{d}^{m}_q\,\mathbf{d}^{h}_q\leq 0$. Considering the Gibbs equation found by Sussman and Larena in \cite{Sussman:2013xpa,Sussman:2015bea} and comparing it with \eref{dotsgrav} we obtain 
\begin{equation}
    \dot{s}_{\textrm{\tiny{(gr)}}}=\frac{T_{\textrm{\tiny{(gr)}}}^{\textrm{\tiny{LTB}}}}{T_{\textrm{\tiny{(gr)}}}}\dot{s}_{\textrm{\tiny{(gr)}}}^{\textrm{\tiny{LTB}}}=\frac{|H_q+3\mathbf{d}_q^H|}{|H_q+3\mathbf{D}_q^H|}\dot{s}_{\textrm{\tiny{(gr)}}}^{\textrm{\tiny{LTB}}},\label{eq:SzekLTBS}
\end{equation}
which reinforces the fact that Szekeres I models inherit the conditions for gravitational entropy growth from their LTB seed models. 
Let us emphasize that Szekeres I models have more degrees of freedom than LTB models, specially on the concavity of density profiles that determine whether a structure is an over-density or a void. As mentioned in \sref{sec:FvsC} and illustrated in figures \ref{Fig:perfini} and \ref{Fig:perfsev}, the LTB seed models used to generate multiple strucures (in particular those used in the numerical example in this paper) do not necessarily have the simple monotonic (pure over-density or void) radial profiles of covariant scalars ($A,\,A_q$) typical of single spherical structures in an FLRW background. Instead, the radial profile of these scalars in the LTB seed model might have more complicated (angle dependent) concavity patterns, suitable to model the elongated ``pancake-like'' structures outside the spheroidal central region that are shown in the figures \ref{Fig:deltas}-\ref{Fig:deltascurvas} (see \cite{Sussman:2015fwa,Sussman:2015wna}).  Considering \eref{Ddrelation} and \eref{eq:SzekLTBS} we can now examine how to generalize properties found for LTB models in \cite{Sussman:2013xpa,Sussman:2015bea} to Szekeres I models. We remark that the material presented in this section can be helpful in dealing with the integrability conditions of $s_{\textrm{\tiny{(gr)}}}$ in \ref{sec:IntCond}.

\subsection{Asymptotic time limits of the gravitational entropy and temperature}

To examine the asymptotic evolution of $\dot{s}_{\textrm{\tiny{(gr)}}}$ and ${s}_{\textrm{\tiny{(gr)}}}$ as $t\to\infty$ we need to find out the time asymptotic forms of the metric functions $a$ and $\mathcal{G}$ from the quadrature of the Friedman-like  equation \eref{adimvars}
\begin{equation} 
H_{qi}(t-t_{bb}) \equiv \mathcal{F}(a) =  \int_1^a{\frac{\sqrt{\hat a}\,d\hat  a}{\left[\Omega_{qi}^m-\Omega_{qi}^k\,\hat a+\bar\Omega_i^\Lambda\,\hat a^3\right]^{1/2}}},\label{intquadrature}
\end{equation} 
where the lower integration limit corresponds to the initial LS time for which the Szekeres I scale factor $a_i=1$ (not to confuse with the background scale factor $\bar{a}$).  Then $\mathcal{G}=\Gamma-\mathbf{W}= 1+ra'/a-\mathbf{W}$ follows from an implicit derivative of $\mathcal{F}$. Since evaluating $\mathcal{F}$ involves complicated elliptic integrals, we use the fact that $\bar\Omega_i^\Lambda\ll 1$ and use \eref{Ddrelation} and \eref{eq:SzekLTBS} in order  to extend to Szekeres I the asymptotic expressions obtained in  the case $\Lambda>0$ in equation (23) of \cite{Sussman:2015bea} for LTB voids in a $\Lambda$CDM background (please notice that we use a different notation from \cite{Sussman:2015bea}: our initial $\Omega$'s are now normalized with $\bar H_i$, not with $H_{qi}$). The result for $a\gg 1$ and $t\gg t_i$ is
\begin{eqnarray}
\fl a\approx \exp\left(\sqrt{\bar\Omega_i^\Lambda}\,H_{qi}\,t\right),\qquad \mathcal{G}\approx 1+3\delta_{qi}^h -\frac{3\left[(\Omega_{qi}^m+\bar\Omega_i^\Lambda)\delta_{qi}^h-\frac12 \Omega_{qi}^m \delta_{qi}^m\right]}{\mathcal{H}_{qi}^2\, a^2}-\mathbf{W},\label{aGasympt}
\end{eqnarray}
which inserted in \eref{eq:flucts} and \eref{adimfluctsdef} yields, by using \eref{Ddrelation} the Szekeres I version of equation (25a) of \cite{Sussman:2015bea} 
\begin{equation} 
\mathbf{D}^m_q\,\mathbf{D}^h_q\approx -\frac{(\delta_{qi}^m-2\delta_{qi}^h)(\delta_{qi}^m-3\delta_{qi}^h)}{\mathcal{H}_{qi}\sqrt{\bar\Omega_i^\Lambda}\,a^5}\,\left(\frac{\Gamma-\mathbf{W}}{\Gamma}\right)^2<0.\label{DDasympt}
\end{equation}
Here we used the approximations $\Omega_{qi}^m\approx 1,\,\mathcal{H}_{qi}\approx 1$ and $\bar\Omega_i^\Lambda\ll 1$ and the negative sign is justified by the signs of the gradients of the initial quasilocal scalars displayed in figure \ref{Fig:perfini}. These show that $\delta_{qi}^m=r[\Omega_{qi}^m]'/(3\Omega_{qi}^m)\geq 0$ and $\delta_{qi}^h=r[\mathcal{H}_{qi}]'/(3\mathcal{H}_{qi})\leq 0$ hold for the LTB seed model used in the numerical example (notice that the sign of $\delta_{qi}^h$ is consistent with $H'_q<0$ because the maximal Hubble expansion occurs at the void center $r=0$ and decreases as $r$ grows). 

The asymptotic form of the gravitational temperature \eref{gravT} also follows from applying the asymptotic forms \eref{aGasympt}. From  \eref{eq:flucts}, \eref{eq:Hconstr} and \eref{eq:DHconstr} we obtain $\mathbf{D}_q^h\sim a^{-2}\to 0$, leading as $a\to\infty$ to :
\begin{equation} 2\pi T_{\textrm{\tiny{(gr)}}} \to \bar H_i \sqrt{\bar\Omega_i^\Lambda},\end{equation}     
which, as mentioned in \cite{Sussman:2015bea}, provides a nice interpretation of the cosmological constant in terms of the asymptotic gravitational temperature in models asymptotic to an FLRW background with $\Lambda>0$.

Following \cite{Sussman:2015bea}, we insert the asymptotic forms \eref{aGasympt} into \eref{eq:flucts}, \eref{eq:Hconstr} and \eref{eq:DHconstr} and substitute in \eref{dotsgrav} to obtain in the asymptotic regime
\begin{equation} 
\dot{\tilde{s}}_{\textrm{\tiny{(gr)}}} \approx \frac{3\bar H_i \mathbf{D}_{qi}^k}{8\bar\Omega_i^\Lambda}\,\exp\left(-2\sqrt{\bar\Omega_i^\Lambda}\,t\right)>0,\label{sdotasympt}
\end{equation}
where  $\tilde{s}_{\textrm{\tiny{(gr)}}}={s}_{\textrm{\tiny{(gr)}}}/s_i$, with $s_i$ an initial value of the gravitational entropy that depends on the spatial coordinates. The positive sign follows from the fact that $\mathbf{D}_{qi}^k>0$, as shown by the radial profile of $\Omega_{qi}^k$ in figure \ref{Fig:perfini}, which is consistent with quasilocal spatial curvature becoming less negative as $r$ increases. The form of \eref{sdotasympt} shows that the gravitational entropy  tends asymptotically to a finite ``saturation'' value obtained from the integral ${s}_{\textrm{\tiny{(gr)}}}|_\infty =s_i +\lim_{a\to\infty}\int_1^a{\dot{{s}}_{\textrm{\tiny{(gr)}}} dt}$. The same saturation limit occurs in LTB models examined in \cite{Sussman:2015bea}, but for Szekeres I models this saturation value depends on the three spatial coordinates. 

For the case $\Lambda=0$ we have in the asymptotic limit $T_{\textrm{\tiny{(gr)}}} \to 0$ and a much slower logarithmic decay $\dot{\tilde{s}}_{\textrm{\tiny{(gr)}}}\sim \ln\,a/a$. Hence, as in \cite{Sussman:2015bea}, we have $\tilde{s}_{\textrm{\tiny{(gr)}}}\to\infty$ regardless of the initial conditions.

\subsection{Gravitational entropy growth vs density growing/decaying modes}

One of the main results in LTB models found in \cite{Sussman:2013xpa,Sussman:2015bea} is the fact that gravitational entropy growth is associated with setting up initial conditions with a growing mode being exclusive or dominant over a nonzero decaying mode. In all examined LTB models $\dot{s}_{\textrm{\tiny{(gr)}}}<0$ occurred when the decaying mode was dominant (in the ``parabolic'' models with $\Omega_q^k=\Lambda=0$ and in cosmic times close to a non-simultaneous Big Bang). It is important to verify whether this property also holds for Szekeres I models describing arrays of multiple structures.

While the growing and decaying modes are a feature of solutions of density perturbations in a linear regime, these concepts can be generalized to exact quantities in the fully non-linear regime of LTB models \cite{Sussman:2013qya}. Such quantities are expressible in terms of elementary functions only when $\Lambda=0$ and in some spacial cases with $\Lambda>0$ (see the explicit elementary functions for $\Lambda=0$ in \cite{Sussman:2013qya}, and also \cite{Sussman:2013xpa}). 

To examine for Szekeres I models the connection between growing/decaying modes and the sign of $\dot{s}_{\textrm{\tiny{(gr)}}}$ that occurs in LTB models, we remark that the cosmological constant has a negligible value at the last scattering initial slice $z=1100$ that we are considering. From \eref{obsparst0} and \eref{obsparsLSS} we have 
\begin{equation}
\bar \Omega_{qi}^\Lambda = \left(\frac{\bar H_0}{\bar H_i}\right)^2\,\bar \Omega_0^\Lambda = 2.38\times 10^{-9}\,\bar \Omega_0^\Lambda\ll 1,
\end{equation}
which justifies using the analytic solutions of the case $\Lambda=0$ to examine the expressions derived in \cite{Sussman:2013qya} for the growing/decaying modes at the initial slice $z=1100$.     

The exact dimensionless density fluctuation $\delta_q^m$ in LTB models satisfies the same second order differential equation \eref{perturb2ord2} as its Szekeres I equivalent $\Delta_q^m$. Since this equation generalizes the evolution equation of linear density spherical dust perturbations over an FLRW background, its exact solution $\delta_q^m$ must generalize the solution $\delta_{\textrm{\tiny{(lin)}}}$ of the second order linear perturbation equation (linear limit of \eref{perturb2ord1}) expressible as the sum of a growing and decaying mode: 
\begin{equation} 
\delta_{\textrm{\tiny{(lin)}}} =  A_{+}(r)\,J_{+}(t)+A_{-}(r)\,J_{-}(t),\label{lindelta}
\end{equation}
where the $A_{\pm}(r)$ are the amplitudes of the growing ($+$) and decaying ($-$) modes $J_{\pm}$. For an LTB model with a spatially flat Einstein de Sitter background, we have in the linear regime $J_{+}\sim t^{2/3}$ and $J_{-}\sim 1/t$, while $A_{-}\propto H_q\,t_{bb}'$, with $t_{bb}(r)$ representing the inhomogeneous Big Bang time. Hence the decaying mode can be suppressed by setting up a simultaneous Big Bang ($t_{bb}'= 0$), but (if shell crossings are absent and initial conditions are set up in a linear regime at early cosmic times) it is dominant already at very early times for a non-simultaneous Big Bang $t_{bb}'\ne 0$. These results can be easily generalized to Szekeres I models with an FLRW background. 

\begin{figure}[ht]
\centering
\begin{subfigure}{.5\textwidth}
  \centering
  \includegraphics[width=0.9\linewidth]{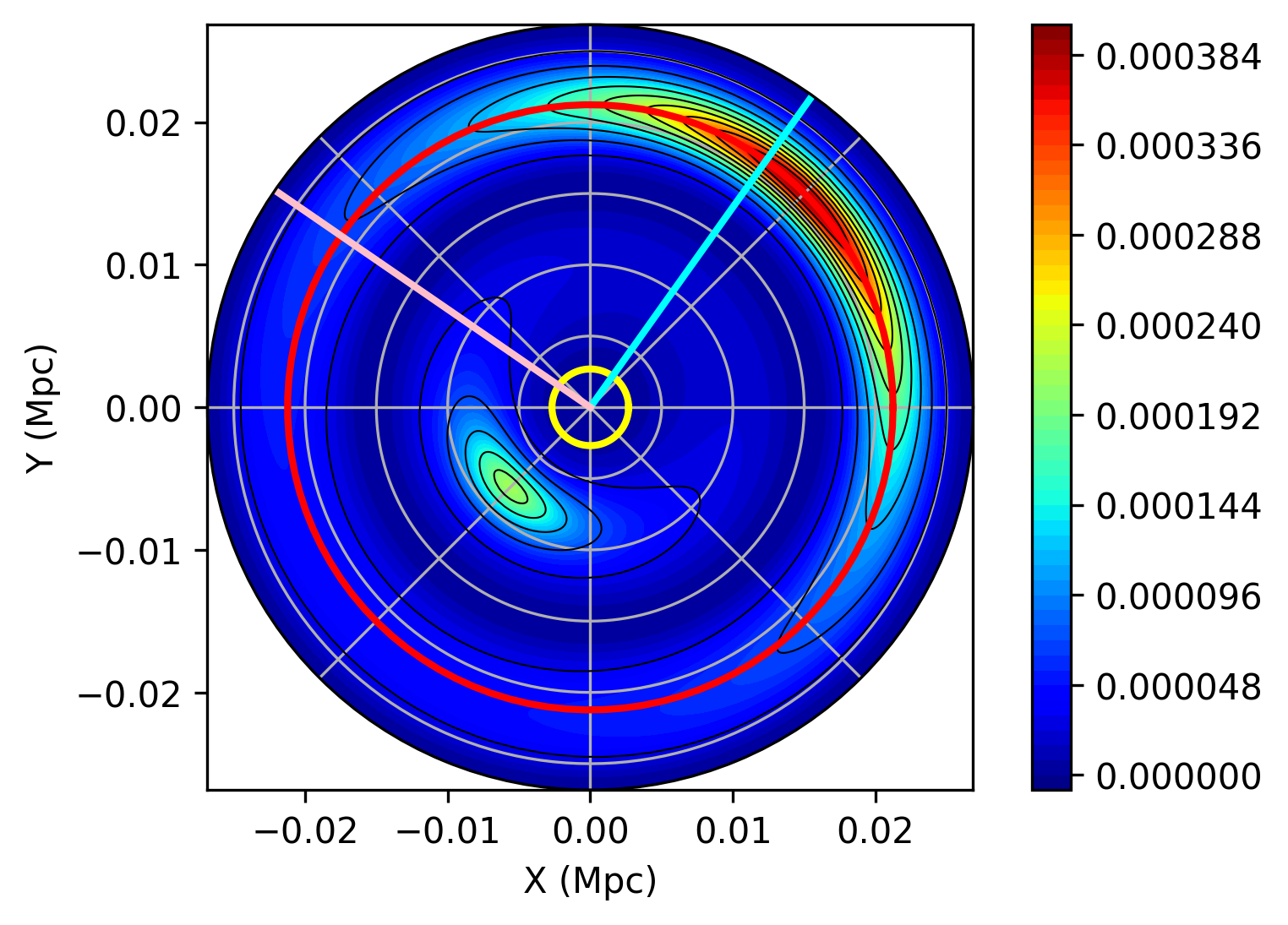}
  \caption{$\Delta^{m}_q$ at $z=1100$.}
  \label{Fig:Deltmi}
\end{subfigure}%
\begin{subfigure}{.5\textwidth}
  \centering
  \includegraphics[width=0.9\linewidth]{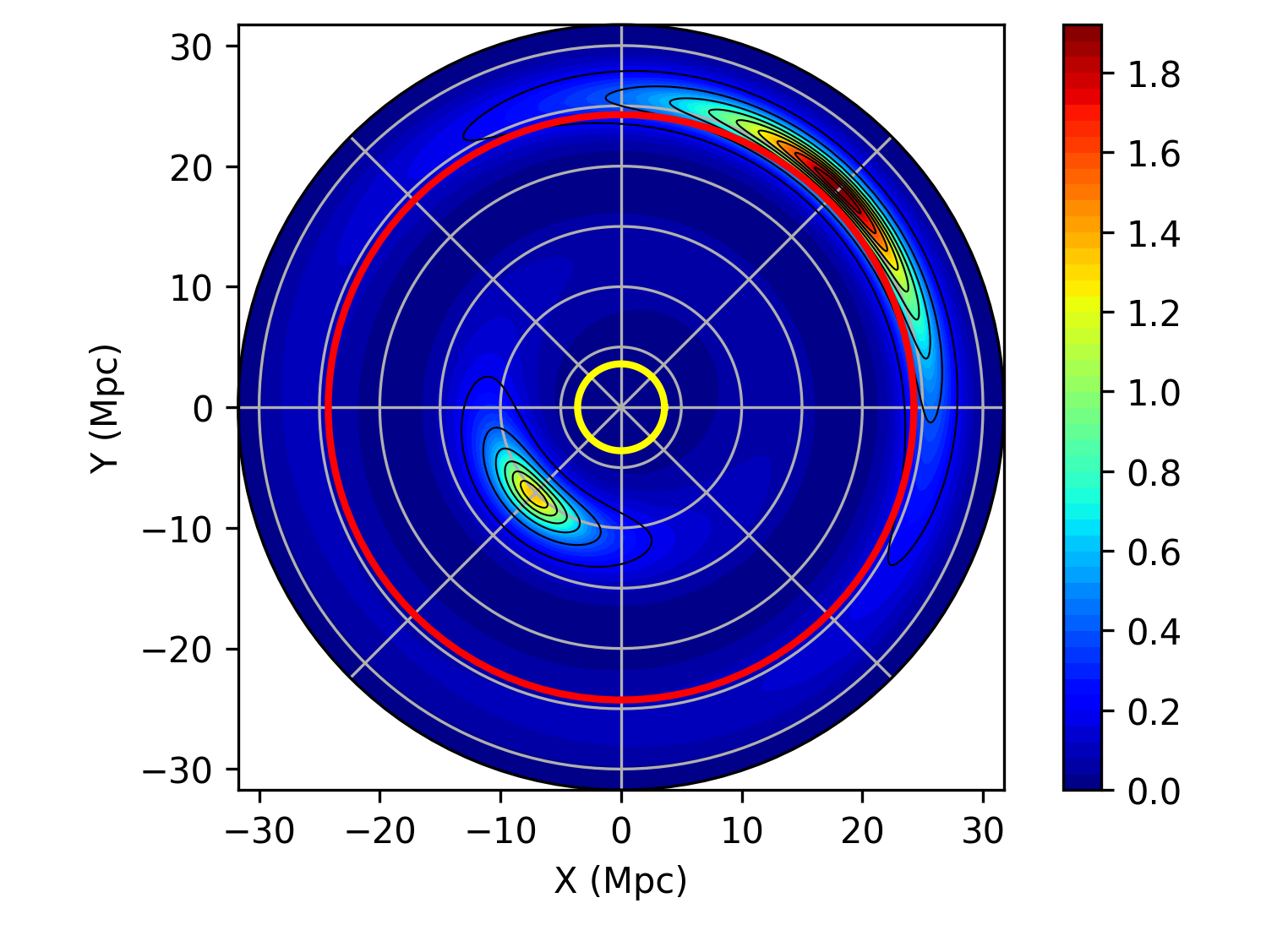}
  \caption{$\Delta^{m}_q$ at $z=0$.}
  \label{Fig:Deltmf}
\end{subfigure}
\begin{subfigure}{.5\textwidth}
  \centering
  \includegraphics[width=0.9\linewidth]{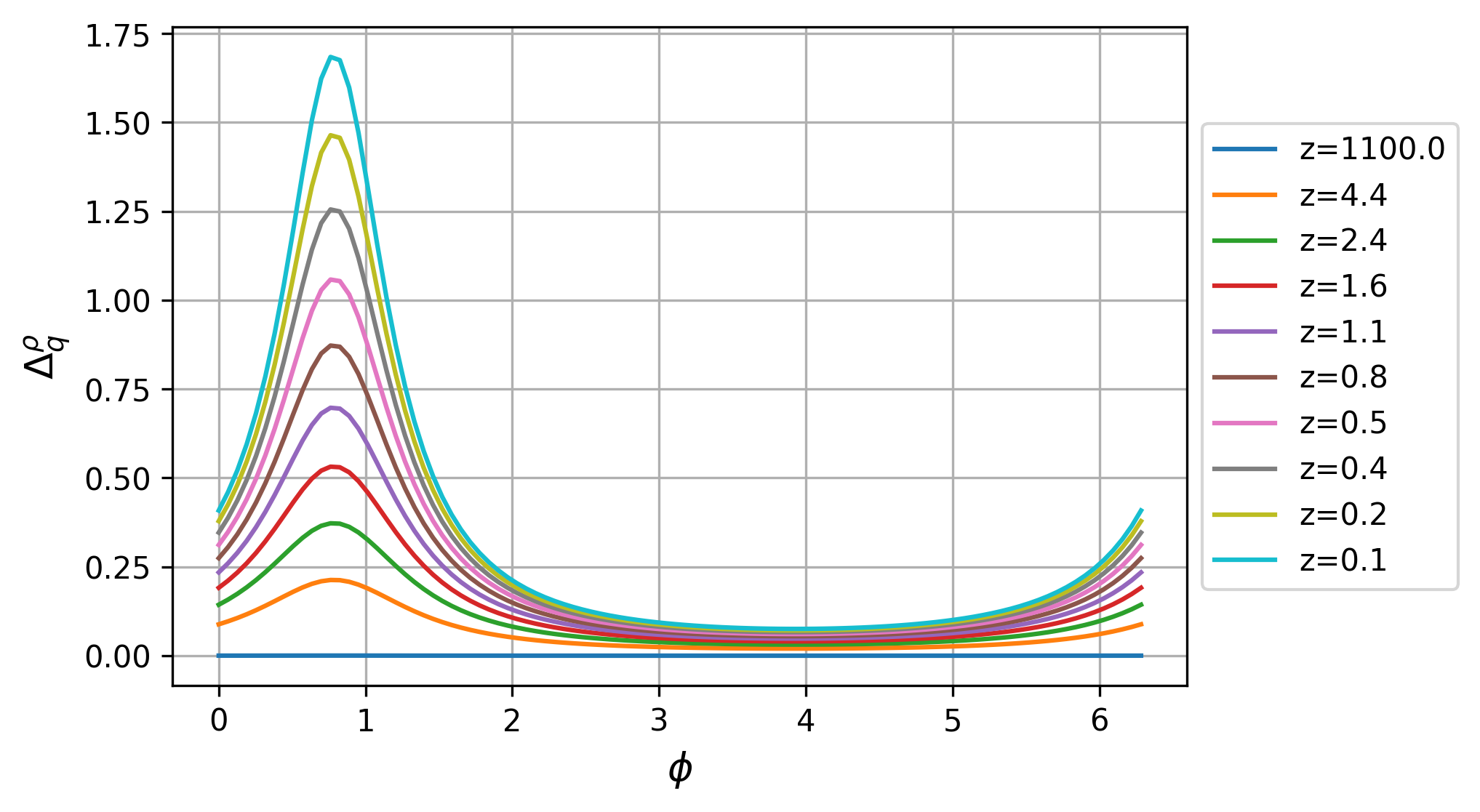}
  \caption{}
  \label{Fig:Deltmphii}
\end{subfigure}%
\begin{subfigure}{.5\textwidth}
  \centering
  \includegraphics[width=0.9\linewidth]{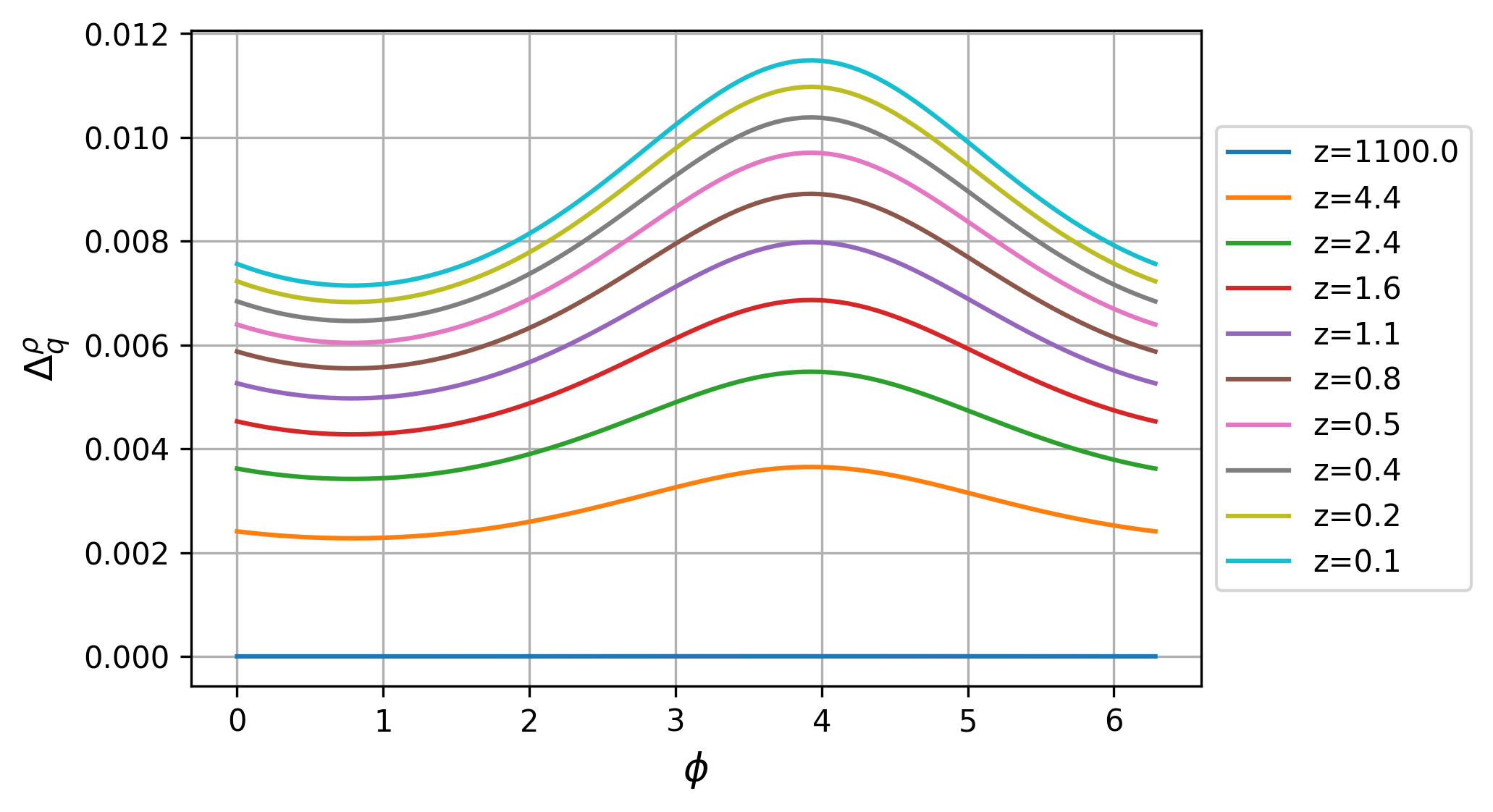}
  \caption{}
  \label{Fig:Deltmphif}
\end{subfigure}
\caption{The figure compares the values of $\Delta^{\rho}_q$ at $z=1100$ (left panels) and $z=0$ (right panels). Panel a shows the radial rays with fixed angles $\phi_1,\,\phi_2$ (respectively passing and avoiding the over-density and with cyan and pink colors). Panels c and d display the curves for $\Delta^{\rho}_q$ as a function of the angle $\phi$ in radians (for fixed $\theta=\pi/2$ and fixed radius passing through an over-density, shown in red in panels a and b). Note that $\Delta^{\rho}_q$ grows with time for all angles (and for all fixed radii, though we do not display this), thus reinforcing the fact that the growing mode is dominant from the initial times onwards.}
\label{Fig:Deltasq}
\end{figure}

The exact generalization of \eref{lindelta} for generic LTB models with $\Lambda=0$ (see equation (23) of \cite{Sussman:2013qya} and equation (14) of \cite{Sussman:2013xpa}) is evidently a nonlinear expression that can be readily extended to Szekeres I models with $\Lambda=0$. From the analytic solutions for the case $\Lambda=0$ in \ref{analytic} we have 
\begin{equation}
\Delta_q^{\rho} = \frac{1+\Delta_{qi}^m-\tilde{\mathcal{G}}}{\tilde{\mathcal{G}}} = \frac{A_{+}\,J_{+}+A_{-}\,J_{-}}{1-A_{+}\,J_{+}-A_{-}\,J_{-}},\label{Deltamodes} \end{equation}
where $\tilde{\mathcal{G}}$ is given by \eref{eq:Gev} and now the amplitudes $A_{\pm}$ are functions of the three spatial coordinates $(r,\theta,\phi)$ (or ($r,x,y$)), while $J_{\pm}$ depend on $(t,r)$ (it reduces to \eref{lindelta} in the linear regime in which $|A_\pm\,J_\pm|\ll 1$ hold). The specific functional forms for negative spatial curvature (as in the numerical example) are
\begin{eqnarray} 
A_{+}&=&\frac{3(\Delta_{qi}^m -\frac32\Delta_{qi}^k)}{1+\Delta_{qi}^m},\qquad  J_{+}= H_q\,(t-t_{bb})-\frac23,\label{gmode}\\
A_{-}&=&\frac{rt'_{bb}}{(1+\Delta_{qi}^m)(1 -\mathbf{W})},\qquad J_{-}=H_q,\label{dmode}  
\end{eqnarray}
with
\begin{eqnarray}
\fl H_q(t-t_{bb}) = \frac{\sqrt{2+\alpha_{qi}\,a}\left[\sqrt{\alpha_{qi}\,a}\sqrt{2+\alpha_{qi}\,a}-\hbox{arccosh}( 1+\alpha_{qi}\,a)\right]}{(\alpha_{qi}\,a)^{3/2}},
\quad \alpha_{qi} = \frac{2|\Omega_{qi}^k|}{\Omega_{qi}^m},\nonumber\\
&{}& \label{Hqttbb}\\
\fl \frac{r t'_{bb}}{3}=\frac{\delta_{qi}^m-\delta_{qi}^k-\left(\delta_{qi}^m-\frac32\delta_{qi}^k\right)H_{qi}(t_i-t_{bb})}{H_{qi}},\label{tbbr}
\end{eqnarray} 
where $H_q$ is given by \eref{eq:DHconstr} with $\Lambda=0$ and $H_q(t-t_{bb})$ follows from the analytic solutions of the case $\Lambda=0$, while the Big Bang time $t_{bb}$ and its gradient $t'_{bb}$ can be obtained from evaluating these solutions at the initial slice in which $a_i=1,\,\Gamma_i=1$ hold (see details in \ref{analytic}).
\begin{figure}[ht]
\centering
\begin{subfigure}{.5\textwidth}
  \centering
  \includegraphics[width=1\linewidth]{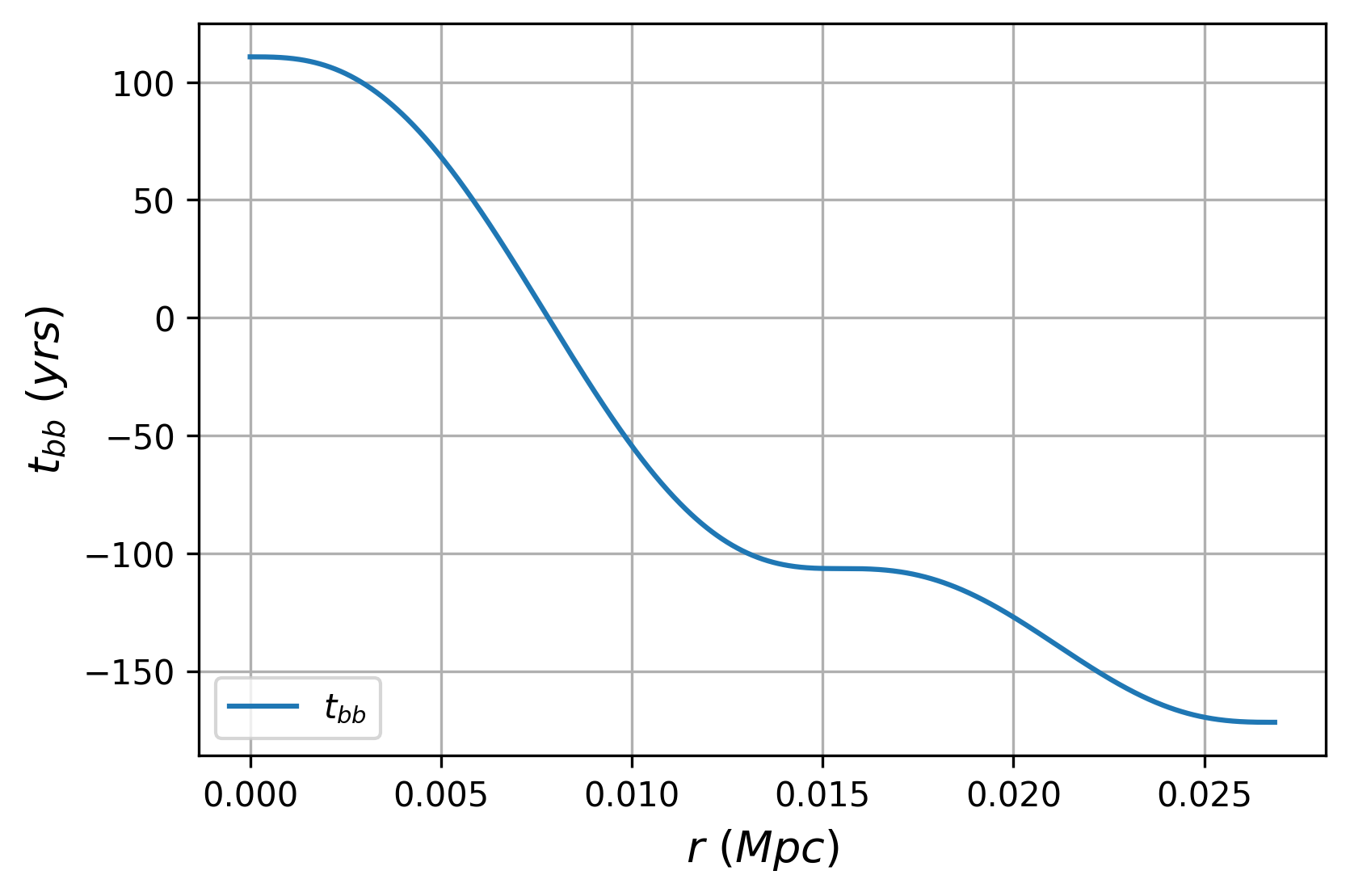}
  \label{Fig:tbb}
\end{subfigure}%
\begin{subfigure}{.5\textwidth}
  \centering
  \includegraphics[width=1\linewidth]{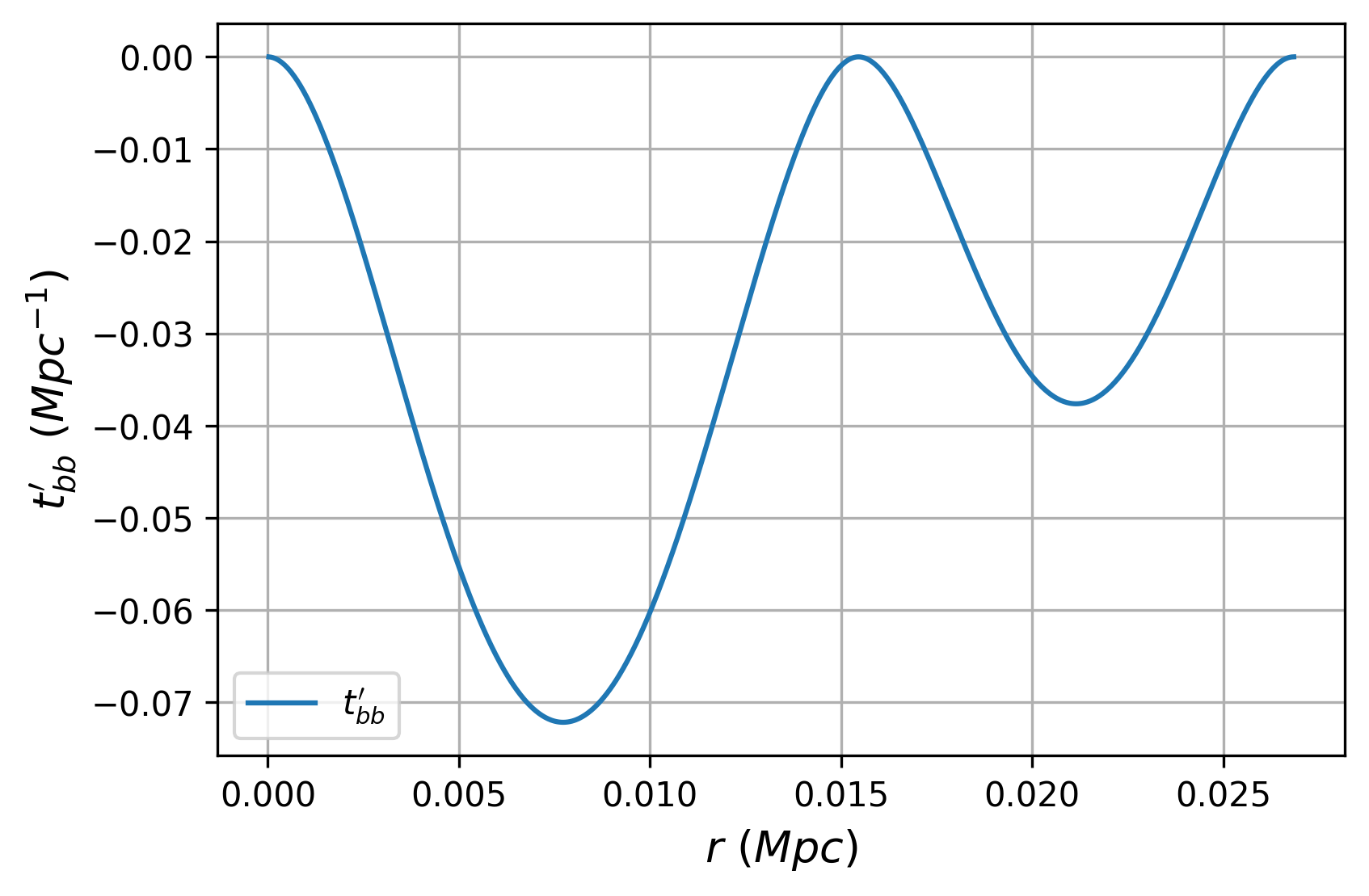}
  \label{Fig:tbbprim}
\end{subfigure}
\caption{The figure displays the Big Bang time $t_{bb}(r)$ (left panel) and its radial gradient $t'_{bb}(r)$ (right panel) respectively obtained from \eref{Hqttbb} and \eref{tbbr}. Notice that $t'_{bb}\leq 0$ holds everywhere, so that $t_{bb}$ decreases with increasing $r$, thus implying a (roughly) 250 years difference in cosmic age between observers at $r=0$ and at the $\Lambda$CDM background, a negligible  difference in comparison with the cosmic age estimates of $\sim 13$ Gys.}
\label{Fig:tbbang}
\end{figure}
As shown in figures \ref{Fig:DmYDh} and \ref{Fig:deltascurvas}, the gravitational entropy is already growing initially since $\mathbf{D}_{qi}^m\mathbf{D}_{qi}^h<0$ holds at $z=1100$. It is straightforward to show by evaluating \eref{gmode}-\eref{dmode} at $t=t_i$ for the initial conditions defined in sections \ref{sec:CondIn}-\ref{multiple} that the growing mode is already dominant: $A_{+}J_{+}|_{t_i}> A_{-}J_{-}|_{t_i}$. Also, as shown in figure \ref{Fig:Deltasq}, $\Delta_q^m$ is rapidly increasing for times $t>t_i$ (or $z<1100$), showing that the growing mode remains dominant with $J_{-}\propto H_q$ rapidly decreasing as the structures evolve and gravitational entropy grows. 

While earlier times with $z \gg 1100$ are not physically relevant because a dust source is not valid for radiation dominated conditions, it is still interesting to examine the behavior of $\dot{s}_{\textrm{\tiny{(gr)}}}$ for times approaching the non-simultaneous Big Bang. From the analytic expressions \eref{Deltamodes}-\eref{tbbr} it follows that as $a\to 0$ the decaying mode becomes dominant: $A_{+}J_{+}\to 0$ while $A_{-}J_{-}\to -\infty$ (since $t'_{bb}<0$ but $H_q\to\infty$). In this limit $\Delta_q^m\to -1$ while $\Delta_q^h = \Gamma/(\Gamma-\mathbf{W})\delta_q^h <0$ (since $\delta_q^h \propto H'_q <0$ in time slices intersecting the Big Bang \cite{Sussman:2010ew}), hence we have $\mathbf{D}_{qi}^m\mathbf{D}_{qi}^h>0$ and thus gravitational entropy decreases: $\dot{s}_{\textrm{\tiny{(gr)}}}<0$, as it happens with LTB models in the same limit close to a non-simultaneous Big Bang.

\section{Conclusions}\label{conclusion}

We have examined the CET proposal of gravitational entropy in the context of Szekeres I models ({\it i.e.} quasispherical models of class I), which are non-trivial inhomogeneous cosmologies that are not constrained by spherical symmetry or other isometries We thus provide an appropriate framework to describe the dynamics of jointly evolving cosmological inhomogeneities from linear initial conditions at last scattering ($z=1100$) towards a fully non-linear regime. 

By means of theoretical arguments and a generic numerical example describing a spheroidal void surrounded by two overdensities, we have shown that for Szekeres I models the CET proposal provides a viable gravitational entropy associated with structure formation as an irreversible process defined by the positive entropy production condition $\dot{s}_{\textrm{\tiny{(gr)}}} > 0$. As shown by the numerical example, this condition precisely holds in the regions where these structures exist and grow. 

By considering the connection with previous work on applying the CET proposal to spherically symmetric LTB models \cite{Sussman:2013xpa,Sussman:2015bea}, we have shown that in spite of their enhanced dynamical freedom, Szekeres I models do inherit two important properties of the CET proposal: (i) entropy production occurs when the exact generalization of the linear growing mode is dominant and (ii) the CET entropy itself reaches a finite position dependent asymptotic value associated with an asymptotic gravitational temperature proportional to $\Lambda>0$. 

We believe that the present article provides the most robust probe so far of the viability of the CET proposal. However, it is clear that Szekeres I models are still limited: by having a dust source (interpreted as CDM) and being Petrov type D, they belong to the class of ``silent'' models in which local observers are unable to exchange information through wavelike effects (either acoustic waves through nonzero non-trivial pressure or gravitational waves by means of the Magnetic Weyl tensor). Also, as commented before, the CET proposal only yields a unique effective energy tensor to derive gravitational entropy for spacetimes that are Petrov types D and N. There is still room for further evaluations since a Szekeres I model of multiple overdensities could collapse and form an apparent horizon as shown in \cite{DelgadoGaspar:2018pmf}. This potentially represents a connection with the more familiar definitions of entropy of a black hole; an issue to be explored in the future. Therefore, it is important to continue testing this (and other) gravitational entropy proposals on spacetimes whose Petrov types are neither D nor N. A good candidate could be the Szekeres models of class II, which under a fully general energy momentum tensor are Petrov type I, and thus can be in principle compatible with more general sources, such as non-comoving dust, magnetic fields or even gravitational waves \cite{Najera:2020jdm}. These extensions of the present articule are currently under consideration.  

\ack

We thank Ismael Delgado Gaspar for useful discussions and support in building the numerical code. FAP acknowledges financial support from the grants program for postgraduate students of CONACYT. JCH acknowledges financial support from program UNAM-PAPIIT Grant IN107521 ``Sector Oscuro y Agujeros Negros Primordiales".

\appendix

\section{\label{sec:coords} Szekeres models in their usual coordinates}

Szekeres I models are usually presented in an LTB-like line element given as follows
\begin{equation}
    \rmd s^2 = - \rmd t^2 +\frac{(R'-R\mathcal{E}'/\mathcal{E})^2}{1+2E} \rmd r^2+\frac{R^2}{\mathcal{E}^2}[\rmd x^2 +  \rmd y^2]\label{eq:LTBLike},
\end{equation}
where $E=E(r)$ and $\mathcal{E}=\mathcal{E}(r,x,y)$ is given by \eref{eq:GammaEdefs}
\begin{equation}
 \mathcal{E} = \frac{S(r)}{2}\left[ \epsilon+\left( \frac{x-P(r)}{S(r)} \right)^2+\left( \frac{y-Q(r)}{S(r)} \right)^2 \right]. 
\end{equation}
while $R=R(t,r)$ satisfies the Friedman-like equation 
\begin{equation}
    \dot{R}^2 = \frac{2M}{R}+2E+\frac{8\pi}{3}\Lambda R^2, \quad \mbox{with} \quad M = M(r),\label{Rdot}
\end{equation}
where $M=M(r)$. A third free function is the Big Bang time $t_{bb}=t_{bb}(r)$ that follows from the integration of \eref{Rdot}:
\begin{equation} 
t-t_{bb}=\int_0^R{\frac{\sqrt{R}\,dR}{\left[2M+2E\,R+\lambda R^3\right]^{1/2}}},\label{quadr}
\end{equation}
where $\lambda = (8\pi/3)\Lambda$. Note that equations \eref{Rdot} amd \eref{quadr} are  equations \eref{eq:DHconstr} and \eref{intquadrature} written in terms of $R,\,M,\,K$. Szekeres models in this representation have five free parameters: the three dipole functions $S,\,P,\,Q$ and any two of the three free functions $M,\,E,\,t_{bb}$, as the third one can be eliminated by a choice of radial coordinate because the metric \eref{eq:LTBLike} is invariant under rescalings $r=r(\hat r)$.   

The metric \eref{eq:LTBLike} takes the FLRW-like form \eref{eq:FLRWlike} by making the radial coordinate choice $R_i=R(t_i,r)=r$ for an arbitrary initial time slice $t=t_i$, as well as the following parametrization:
\begin{equation}
a = \frac{R}{r}, \qquad \Gamma =\frac{rR'}{R}, \qquad K_{qi}r^2=-2E,\qquad \mathbf{W} = \frac{r\mathcal{E}'}{\mathcal{E}},   
\end{equation}
leading to the density, curvature and expansion q-scalars 
\begin{equation}
    \frac{8\pi}{3}\rho_q = \frac{2M}{R^3}, \quad H_q =\frac{\dot{R}}{R}, \quad K_q = -\frac{2E}{R^2}
\end{equation}
The fluctuations $\mathbf{D}^A_q$ and the scalars $A=\rho, H, \mathcal{K}$ can be computed from \eref{eq:locscals} and \eref{eq:shearEWeyl}. The initial value functions follow from the obtained expressions by setting $R_i=r$ and $R'_i=1$. The non-diagonal metric in stereographic spherical coordinates is then obtained by the transformation \eref{eq:stereogr}.

\section{\label{sec:IntCond}Integrability of the Gibbs 1-form}

As argued in \cite{Sussman:2013xpa, Sussman:2015bea}, $\dot{s}_{\textrm{\tiny{(gr)}}}$ is the projection along $u^a$ of the Gibbs 1-form with integrating factor $T_{\textrm{\tiny{(gr)}}}$
\begin{equation}
    \rmd s_{\textrm{\tiny{(gr)}}}=\frac{\rmd (\rho_{\textrm{\tiny{(gr)}}})}{T_{\textrm{\tiny{(gr)}}}}.
\end{equation}
To know if it is a closed form, we need to explore if the second partial derivatives commute, i.e., $\partial_a\partial_b{s}_{\textrm{\tiny{(gr)}}}=\partial_b\partial_as_{\textrm{\tiny{(gr)}}}$ with $a \neq b$ (in more rigorous terms to verify if $\rmd (\rmd s_{\textrm{\tiny{(gr)}}})=0$ holds). This translates for a general Szekeres I model into
\begin{equation} \partial_{[a} T_{\textrm{\tiny{(gr)}}}\partial_{b]} s_{\textrm{\tiny{(gr)}}}= 0,\label{condd2s} \end{equation}
where the square brackets denote anti-symmetrization in the indices $a,\,b$. For quasi-spherical Szekeres I models equations \eref{condd2s} reduce to a single condition which expressed in terms of $\dot{s}_{\textrm{\tiny{(gr)}}}^{\textrm{\tiny{LTB}}}$ becomes
 \begin{equation}
     \fl \left( \frac{T^{\textrm{\tiny{LTB}}}_{\textrm{\tiny{(gr)}}}}{T_{\textrm{\tiny{(gr)}}}}\right)\partial_rT_{\textrm{\tiny{(gr)}}}\partial_ts_{\textrm{\tiny{(gr)}}}^{\textrm{\tiny{LTB}}}-\left( \frac{T^{\textrm{\tiny{LTB}}}_{\textrm{\tiny{(gr)}}}}{T_{\textrm{\tiny{(gr)}}}}\right)\partial_tT_{\textrm{\tiny{(gr)}}}\partial_rs_{\textrm{\tiny{(gr)}}}^{\textrm{\tiny{LTB}}}=0 \Longrightarrow (T^{\textrm{\tiny{LTB}}}_{\textrm{\tiny{(gr)}}})'\dot{s}^{\textrm{\tiny{LTB}}}_{\textrm{\tiny{(gr)}}}-\dot{T}^{\textrm{\tiny{LTB}}}_{\textrm{\tiny{(gr)}}}(s^{\textrm{\tiny{LTB}}}_{\textrm{\tiny{(gr)}}})' = 0.
 \end{equation}
which as shown in \cite{Sussman:2013xpa} and \cite{Sussman:2015bea} does not hold in general, implying that the 1-form $\rmd s_{\textrm{\tiny{(gr)}}}$ is not closed and thus its integration is path dependent and must be computed through a line integral (see detail in \cite{Sussman:2015bea}). Given the relation between entropy production along worldlines of fundamental observers in Szekeres I and LTB models in equation \eref{eq:SzekLTBS}, it is evident that equations \eref{condd2s} also fail to hold for quasi-spherical Szekeres I models, since the derivatives with respect to $x,\,y$  act on the ratio of gravitational temperatures that has a very specific dependence on these variables (in $\mathbf{D}_q^h$). 

The line integral used in \cite{Sussman:2015bea} acted along a path made by a radial ray (curve parametrized by $r$ with $t,\theta,\phi$ constant) from the symmetry center at $r=0$ to some $r$ and then integrating along a comoving worldline.  Because of spherical symmetry, all radial rays are equivalent (they are geodesics of the time slices), hence the radial profile of $s_{\textrm{\tiny{(gr)}}}$ is sufficient to specify an initial entropy through this line integral. For the non-spherical Szekeres I models this is a much more complicated process that needs to be carefully constructed.  

\section{Analytic solutions for $\Lambda=0$} \label{analytic}

Analytic solutions of the Friedman-like equation with negative spatial curvature $K_{qi}<0$ of quasi-spherical Szekeres I models with $\Lambda=0$ are well known. From section 7.2 of \cite{Sussman:2011bp}  (with $m_{q0}$ and $|\mathcal{K}_{q0}|$ given by $(1/2)\Omega_{qi}^m \bar H_i^2$ and $|\Omega_{qi}^k| \bar H_i^2$) we have 
\begin{eqnarray}
H_{qi}(t-t_{bb}) = \frac{1}{\beta_{qi}}\left[\sqrt{\alpha_{qi}\,a}\sqrt{2+\alpha_{qi}\,a}-\hbox{arccosh}\left(1+\alpha_{qi}\,a\right)\right],\label{analytsols1}\\
\alpha_{qi} = \frac{2|\Omega_{qi}^k|}{\Omega_{qi}^m},\qquad \beta_{qi}=\frac{2|\Omega_{qi}^k|^{3/2}}{\Omega_{qi}^m}.\label{analytsols2}
\end{eqnarray}
Equation \eref{Hqttbb} follows by multiplying \eref{analytsols1} by $\mathcal{H}_q = \sqrt{\Omega_{qi}^m/a^3+|\Omega_{qi}^k|/a^2}$ expressed in terms of $\alpha_{qi}$. The Big Bang time $t_{bb}$ follows by evaluating \eref{analytsols1} at $t=t_i$ so that $a_i=1$, while $t'_{bb}$ in \eref{tbbr} is obtained by its radial derivative using \eref{eq:flucts} to eliminate initial density and curvature gradients in terms of initial fluctuations $\Delta_{qi}^m,\,\Delta_{qi}^k$. The function $\Gamma$ needed to compute \eref{Deltamodes} is also obtained by the radial derivative of \eref{analytsols1} (see equations (58b) and (59) of \cite{Sussman:2011bp}).

 \end{document}